%% file: main.tex
\documentclass[9pt, a4paper, twocolumn]{article}
\input{packages}

\title{The Potential of Subsampling and Inpainting for \\ Fast Low-Dose Cryo FIB-SEM Imaging and Tomography}

\author[1]{D. Nicholls}
\author[2,3]{M. Kobylynska}
\author[4]{J. Wells}
\author[1]{Z. Broad}
\author[1]{A. W. Robinson}
\author[5]{D. McGrouther}
\author[1,6]{A. Moshtaghpour}
\author[6,7]{A. I. Kirkland}
\author[2,3]{ R. A. Fleck} 
\author[1,8,9]{N. D. Browning}

\affil[1]{\normalsize Department of Mechanical, Materials and Aerospace Engineering, University of Liverpool, UK.}
\affil[2]{Centre for Ultrastructural Imaging, King's College London, London, UK.}
\affil[3]{Randall Centre for Cell and Molecular Biophysics, King's College London, London, UK.}
\affil[4]{Distributed Algorithms Centre for Doctoral Training, University of Liverpool, UK.}
\affil[5]{JEOL (UK) Ltd, UK.}
\affil[6]{Rosalind Franklin Instititute, Harwell Science and Innovation Campus, Didcot, UK.}
\affil[7]{Department of Materials, University of Oxford, Oxford, UK.}
\affil[8]{Physical and Computational Science Directorate, Pacific Northwest National Laboratory, Richland, USA.}
\affil[9]{Sivananthan Laboratories, 590 Territorial Drive, Bolingbrook, IL, USA.}
\date{}

\begin{document}

\maketitle

\begin{abstract}
    Traditional image acquisition for cryo focused ion-beam scanning electron microscopy tomography often sees thousands of images being captured over a period of many hours, with immense data sets being produced. When imaging beam sensitive materials, these images are often compromised by additional constraints related to beam damage and the devitrification of the material during imaging, which renders data acquisition both costly and unreliable. Subsampling and inpainting are proposed as solutions for both of these aspects, allowing fast and low-dose imaging to take place in the FIB-SEM without an appreciable loss in image quality. In this work, experimental data is presented which validates subsampling and inpainting as a useful tool for convenient and reliable data acquisition in a FIB-SEM, with new methods of handling 3-dimensional data being employed in context of dictionary learning and inpainting algorithms using a newly developed microscope control software and data recovery algorithm. 
\end{abstract}

\section{Introduction}

\begin{figure*}[ht]
    \centering
    \includegraphics[width=0.55\linewidth]{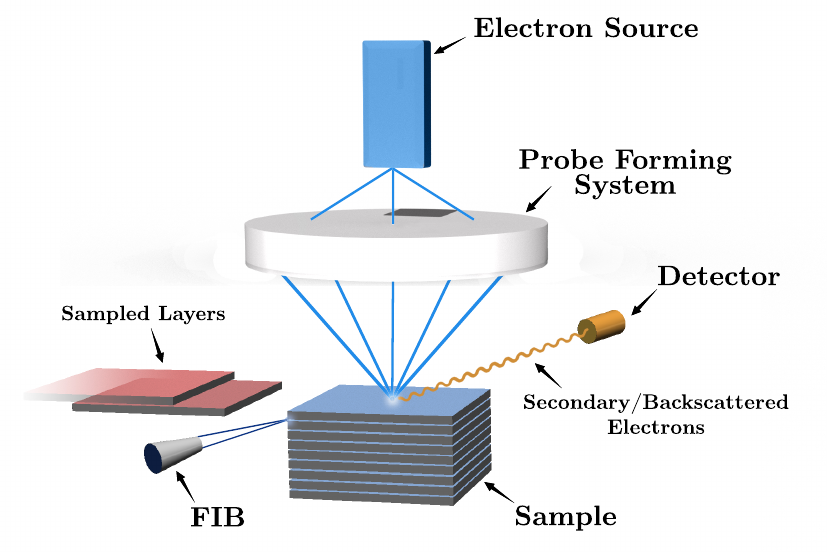}
    \includegraphics[width=0.3\linewidth]{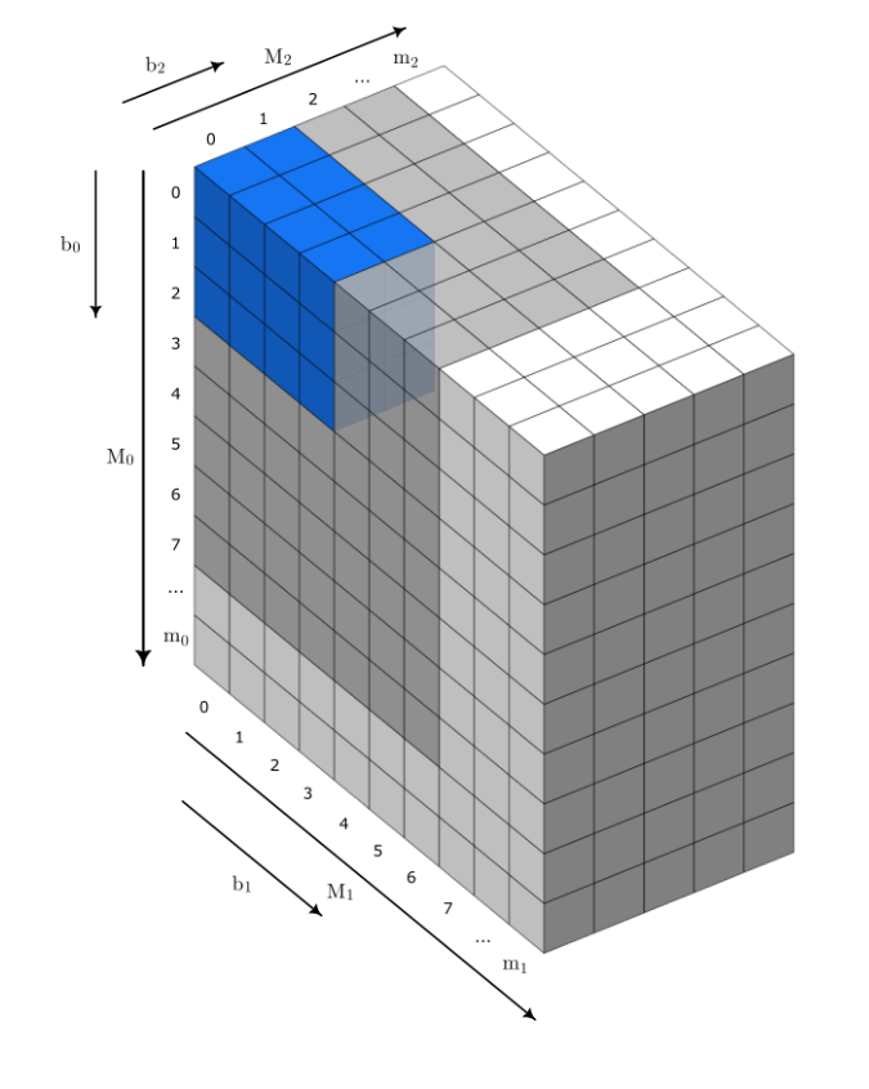}
    \caption{(Left) Operating principles of Cryo FIB-SEM. A focused ion-beam is used to remove a layer of material. A scanning electron microscope is then used to obtain surface information from that newly revealed surface. (Right) Diagram showing the 3-dimensional nature of the data cube and the structure of a patch in 3D for dictionary leaning and inpainting. Each voxel in this diagram is equivalent to a pixel in each image, where the cube is constructed from a series of images. }  
    \label{fig:fib-sem-scheme}
    \vspace{-2mm}
\end{figure*}

Focused ion-beam scanning electron microscopy (FIB-SEM) tomography is a powerful technique for performing high-resolution volume imaging. This technique produces three/four-dimensional SEM image data cubes generated by sequential SEM imaging and FIB serial sectioning [\cite{giannuzzi2004introduction, kizilyaprak2014fib}]. In the case of three-dimensional data, each FIB-section is followed by a single image. In the case of four-dimensional data, each FIB-section is followed by a set of images, referred to individually as frames. The dimensionality of this 4D data can be naturally reduced to 3D by means of integrating each stack of frames, and this is commonly done during the imaging of beam/charge sensitive materials such as those often experienced when employing cryo techniques for the imaging of biological materials. It is observed that acquiring multiple images with a reduced dwell time (\textit{i.e.,} the beam exposure time for each pixel in an image) produces higher quality images when compared to acquiring an equivalent electron-count image at a higher dwell time but with fewer frames. 
The mechanism behind this phenomena is not entirely understood by the community and work related to this is ongoing, but is generally attributed to sample charging. 

For the aforementioned volume imaging of biological materials at cryo conditions [\cite{schertel2013cryo, fleck2015low, fleck2019brief, hayles2021introduction}], there is a constant struggle - the sample must be adequately irradiated to produce enough signal for fine features to be resolved by the scanning electron microscope, yet not too irradiated to induce charging, which affects signal-to-noise (SNR), or worse yet, alter the structure of the material or devitrify it. Greater understanding of sample preparation and more sensitive detectors have made significant strides in lowering the barrier of entry to perform cryo FIB-SEM imaging and tomography, but the process of acquiring significant sized volumes with sufficient quality for performing post-acquisition analysis remains difficult, and requires an incredible time and expertise commitment. It is not uncommon for FIB-SEM tomography data sets to span multiple days or weeks, depending on the imaging conditions and experiment requirements, with regular monitoring. A majority of this time is spent imaging, wherein thousands of low-dose images may be acquired for each frame, before being stitched together to form a full volume. 

It has been previously proposed that compressive sensing fundamentals can be applied to this imaging domain through the application of subsampling and inpainting methods [\cite{nicholls2023targeted}]. This work validated, via simulation, the use of subsampling: the deliberate acquisition of incomplete data sets, and inpainting, the recovery of missing sections of images. This work indicated that taking advantage of the unique dimensionality of the data provided by the FIB-SEM acquisition model is incredibly useful in increasing data acquisition and inpainting efficiency. Investigated here is an extension of this line of enquiry not only with experimental data, but also with a new method for efficiently inpainting data produced by cryo FIB-SEM imaging of a biological system, \textit{Euglena gracilis}.

\section{Methods}

\subsection{Experimental Setup}

All of the data presented in this work was acquired using a JEOL JIB-4700F Z FIB-SEM (JEOL, Japan), equipped with a Leica microsystems EM VCT500 cryo stage and cryo transfer system (Leica microsystems, Austria). The data acquisition was performed using SenseAI's control software (SenseAI, UK) and Quantum Detectors' scan engine (Quantum Detectors, UK). 

The sample, \textit{Euglena gracilis} (4$\mu$L), Klebs CCAP 1224/5Z, was pipetted onto a gold TEM grid with a holey carbon support film (R 1.2/1.3) with 1.2$\mu$m hole diameter, 1.3$\mu$m spacing, and 2.5$\mu$m periodicity (Quantifoil, Germany). Samples were blotted (60s at 98\% RH) and plunge frozen into liquid ethane (EM GP, plunge freezer, Leica microsystems). Vitrified grids were clipped onto a JEOL cryoARM transfer cartridge (JEOL, Japan) and loaded onto a EM VCT transfer block for transfer to the FIB-SEM. Loading was performed at 0.1\% RH in a EM VCM cryo loading station (Leica microsystems) and carried to the FIB-SEM under vacuum in a cryogenically cooled EM VCT500 transfer shuttle. An intermediate evacuation of the EM VCT500 shuttle was applied by attaching the shuttle to a EM VCT500 compatible EM ACE 900 vacuum coating instrument (Leica microsystems). A single \textit{E. gracilis} was identified and an organoplatinum coating (5s) was applied to the surface of the sample to minimise curtaining and aid sample conductivity. A clean face of the specimen was prepared prior to imaging using FIB milling and polishing. This sample was chosen as it is a well understood sample that is well documented, and is a typical proxy for many relevant biological materials.

\begin{figure*}
    \centering
    \begin{minipage}{0.49\textwidth}
        \centering
        \includegraphics[width=\linewidth]{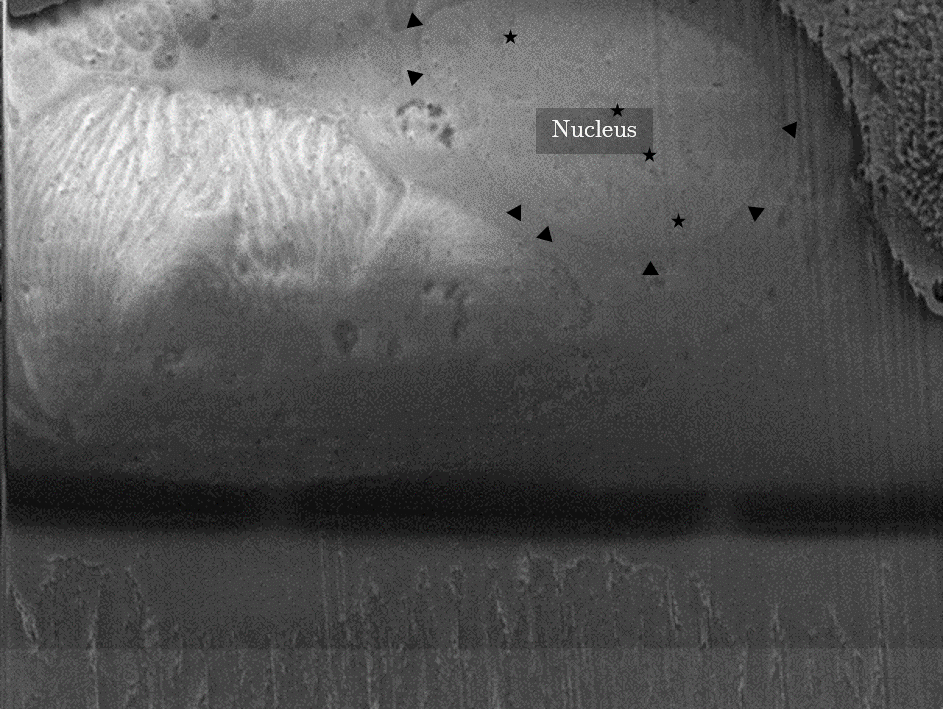}
        \caption*{Fig.~\ref{fig:senseai_example} (a): Nucleus surrounded by nuclear membrane ($\blacktriangle$) and containing heterochromatin ($\bigstar$).}
    \vspace{0.5cm}
    \end{minipage}
    \hfill
    \begin{minipage}{0.49\textwidth}
        \includegraphics[width=\linewidth]{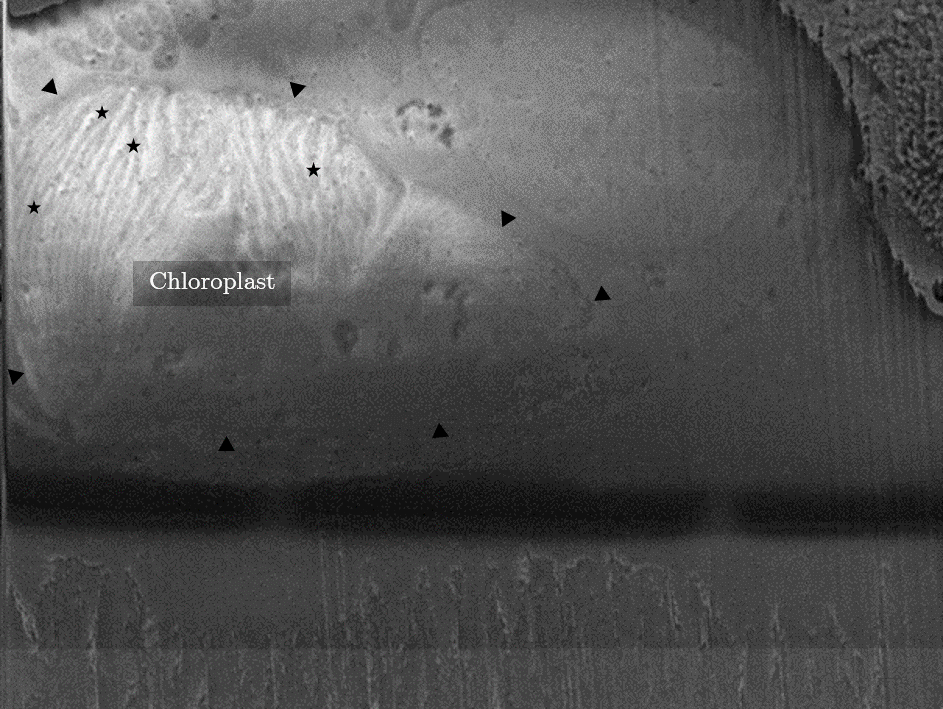}
        \caption*{Fig.~\ref{fig:senseai_example} (b): Chloroplast surrounded by chloroplast membrane ($\blacktriangle$) and containing thylakoid membrane ($\bigstar$).}
    \end{minipage}
    \begin{minipage}{0.49\textwidth}
        \includegraphics[width=\linewidth]{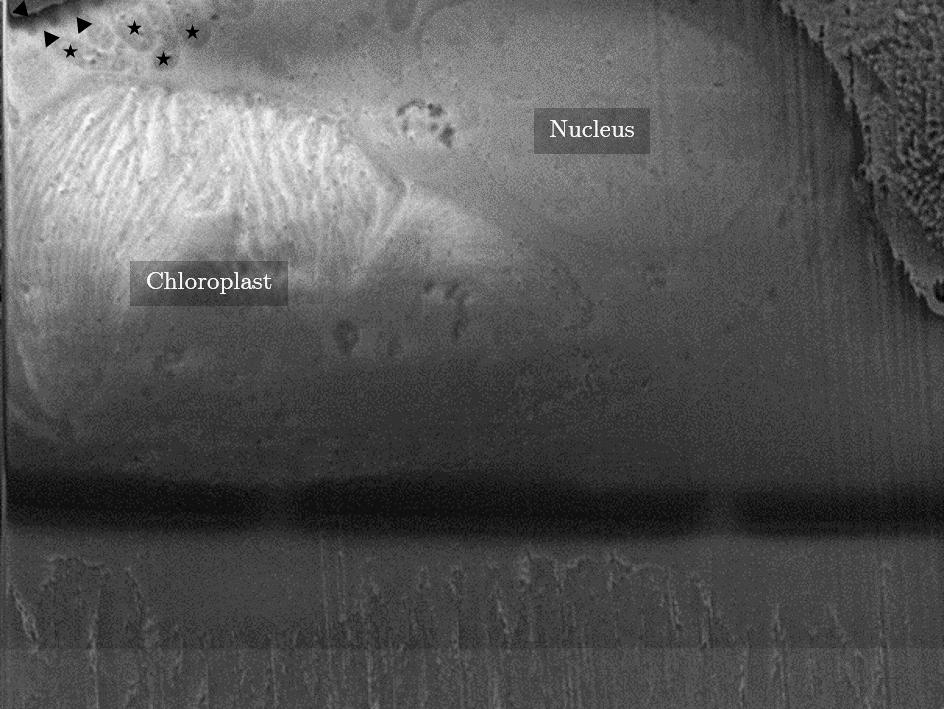}
        \caption*{Fig.~\ref{fig:senseai_example} (c): Mitocondria ($\bigstar$) and outer pellicle structure consisting of a proteinaceous layer or 'membrane skeleton' that is underlain by microtubules and covered by the plasma membrane of the cell ($\blacktriangle$).}
    \end{minipage}
    \caption{SEM secondary electron integrated image formed by SenseAI's microscope control software and Quantum Detectors' scan engine; 100\% sampling, 1$\mu$s dwell time, probe 3, and 100 frames with no processing. Critical structures which are resolved within the image are labelled. Flyback distortions are present in the image (left hand edge) as no flyback compensation is performed. This is done to minimise dose exposure and minimise imaging time. Image dimensions are 21.50 $\times$ 16.12 $\mu$m.} \vspace{-0.3cm}
    \label{fig:senseai_example}
\end{figure*}

To perform the subsampled data acquisition and inpainting, SenseAI's control software and Quantum Detectors' scan engine were utilised. SenseAI is a software suite which provides microscope control to perform regular and subsampled imaging with high levels of user control, as well as performing image reconstruction directly through the use of dictionary learning based image inpainting algorithms. Quantum Detectors' scan engine is used to interface directly with the SEM scanning system, circumventing the manufacturer's image acquisition software. Together, these tools provide the microscopist with powerful options for tailoring their image acquisition to their exact specifications. Fig.~\ref{fig:senseai_example} shows an example of an SEM secondary electron image formed by integration of 100 frames acquired with a 1$\mu$s dwell time. A detailed description of the acquisition model can be found in previous work [\cite{nicholls2023targeted}]. To ensure hysteresis issues are minimised, as they are not the focus of this study, line hop sampling was utilised [\cite{kovarik2016implementing, nicholls2021subsampled}]. Line hop sampling has been previously been implemented on scanning transmission electron microscopes with positive results, and as the optics and system are similar, it was proposed that line hop sampling would be adequate for use with scanning electron microscopes. A detailed study into scanning electron microscope hysteresis when subsampling is reserved for future work, though positive results have been published previously [\cite{anderson2013sparse}].

\subsection{Data Recovery}

For a detailed description of the image recovery model, the reader is referred to previous work [\cite{nicholls2023targeted}] which covers this in depth. The only changes to the image recovery model for this work is that the partitioning of the data cube into $B \times B$ overlapping square patches is instead partitioned into $B \times B \times L$ patches, where $L$ is depth of the patch in the Z axis, through the data cube, and as such the respective dictionary dimensions change to reflect this. 

As previously mentioned, the data was acquired and then inpainted using SenseAI, a microscope control and image recovery software. SenseAI employs the beta process factor analysis (BPFA) algorithm [\cite{paisley2009nonparametric, paisley2014bayesian}], a dictionary leaning algorithm, which has been previously validate for use in other electron microscopy applications [\cite{stevens2014potential, kovarik2016implementing, nicholls2021subsampled, nicholls2022compressive, robinson2022compressed, robinson2022sim2}]. BPFA can be used to learn features within a data set, and when paired with an appropriate sparse coding algorithm, can be used to inpaint data - fill in the missing gaps. By using a 3-dimensional patch shape within SenseAI, information can be learned and inpainted in all three dimensions. This allows dictionary elements to be formed which consider the whole data stack, allowing information from different layers within the stack to inform the learning process. For the examples provided in this work, no image registration or alignment was performed - the data is untreated and used as it is acquired from the microscope. Performance could theoretically be improved by pre-processing, but this is omitted to allow the proposed method to be studied in isolation. 

\section{Results}

\begin{figure*}
    \def\firstColWidth{0.1}
    \def\colWidth{0.215}
    \def\vertSpacing{0.25mm}

    \centering

    \begin{minipage}{\textwidth}
        \centering
        \begin{minipage}{\firstColWidth\textwidth}
            \centering
            Sampling Percentage
        \end{minipage}
        \begin{minipage}{\colWidth\textwidth}
            \centering
            \textbf{Sample Acquisition} \\ (Frame 50)  
        \end{minipage}
        \begin{minipage}{\colWidth\textwidth}
            \centering
            \textbf{Sample Reconstruction} \\ (Frame 50)  
        \end{minipage}
        \begin{minipage}{\colWidth\textwidth}
            \centering
            \textbf{Integrated Acquisition} \\ (100 Frames)  
        \end{minipage}
        \begin{minipage}{\colWidth\textwidth}
            \centering
            \textbf{Integrated Reconstruction} \\ (100 Frames)
        \end{minipage}
    \end{minipage}

    \vspace{\vertSpacing}
    
    \begin{minipage}{\textwidth}
        \centering
 
        \begin{minipage}{\firstColWidth\textwidth}
            \begin{center}
                100\%
            \end{center}
        \end{minipage}
        \begin{minipage}{\colWidth\textwidth}
            \centering
            \includegraphics[width=\textwidth]{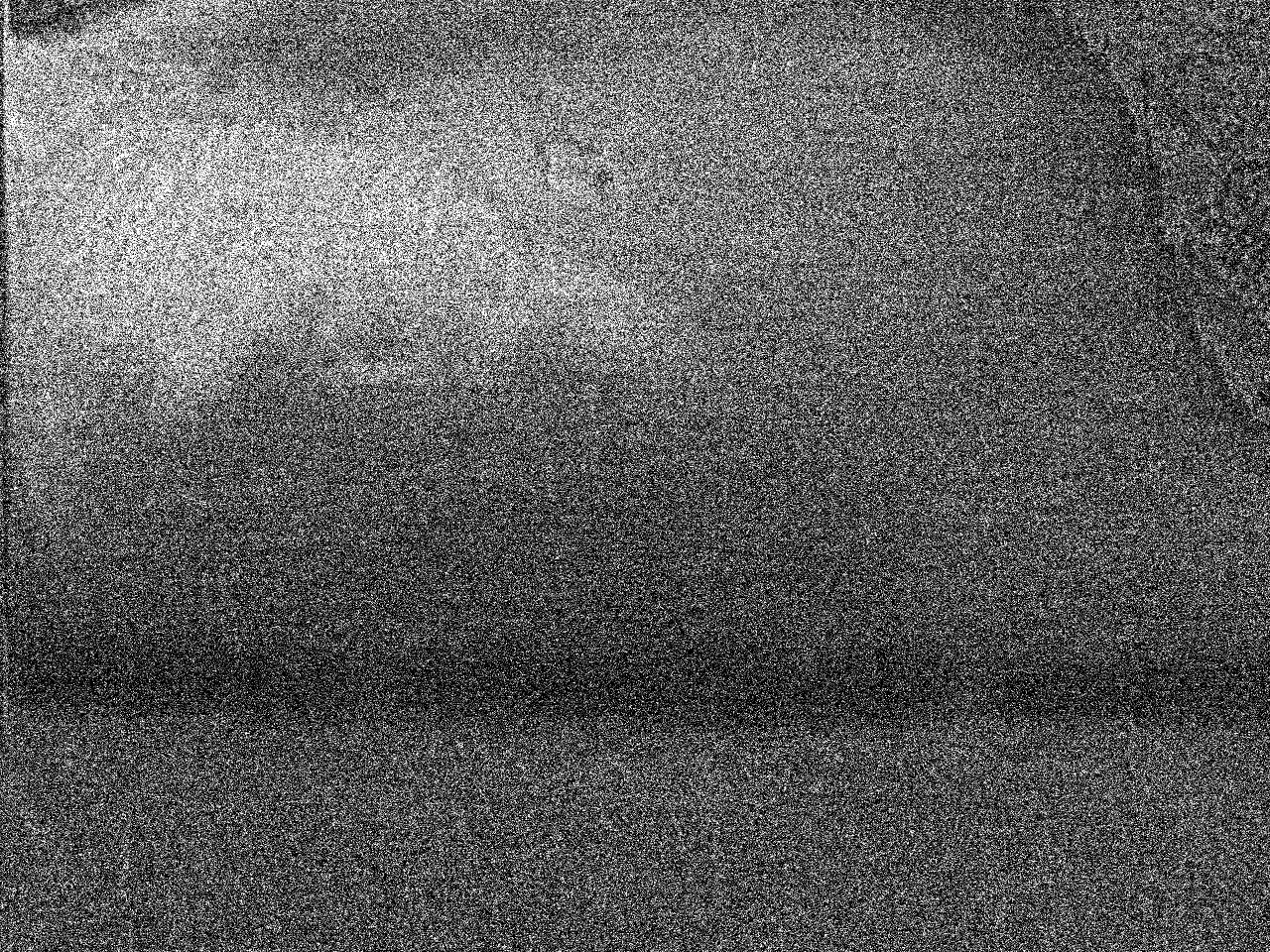}
        \end{minipage}        
        \begin{minipage}{\colWidth\textwidth}
            \centering
            \includegraphics[width=\textwidth]{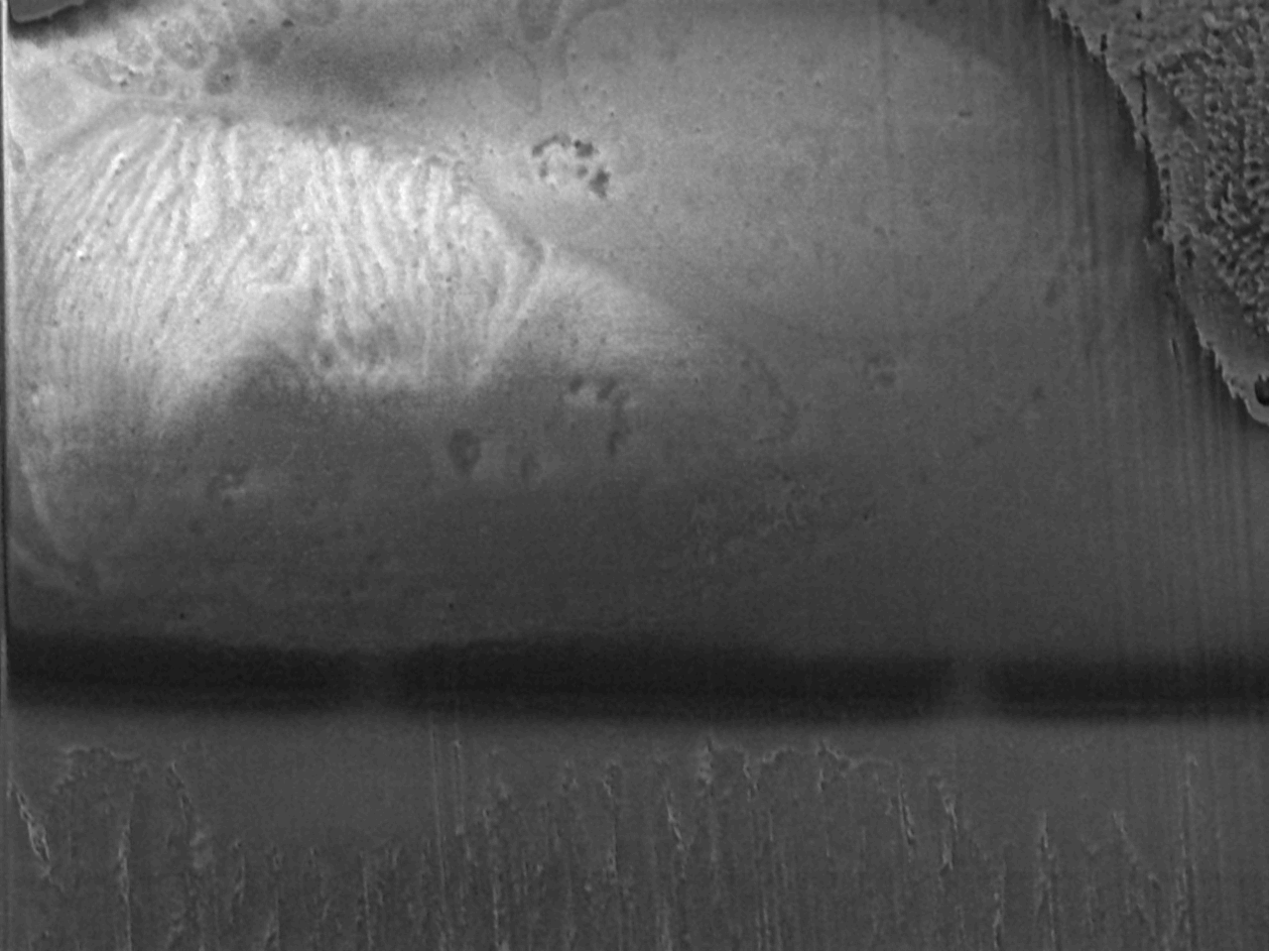}
        \end{minipage}
        \begin{minipage}{\colWidth\textwidth}
            \centering
            \includegraphics[width=\textwidth]{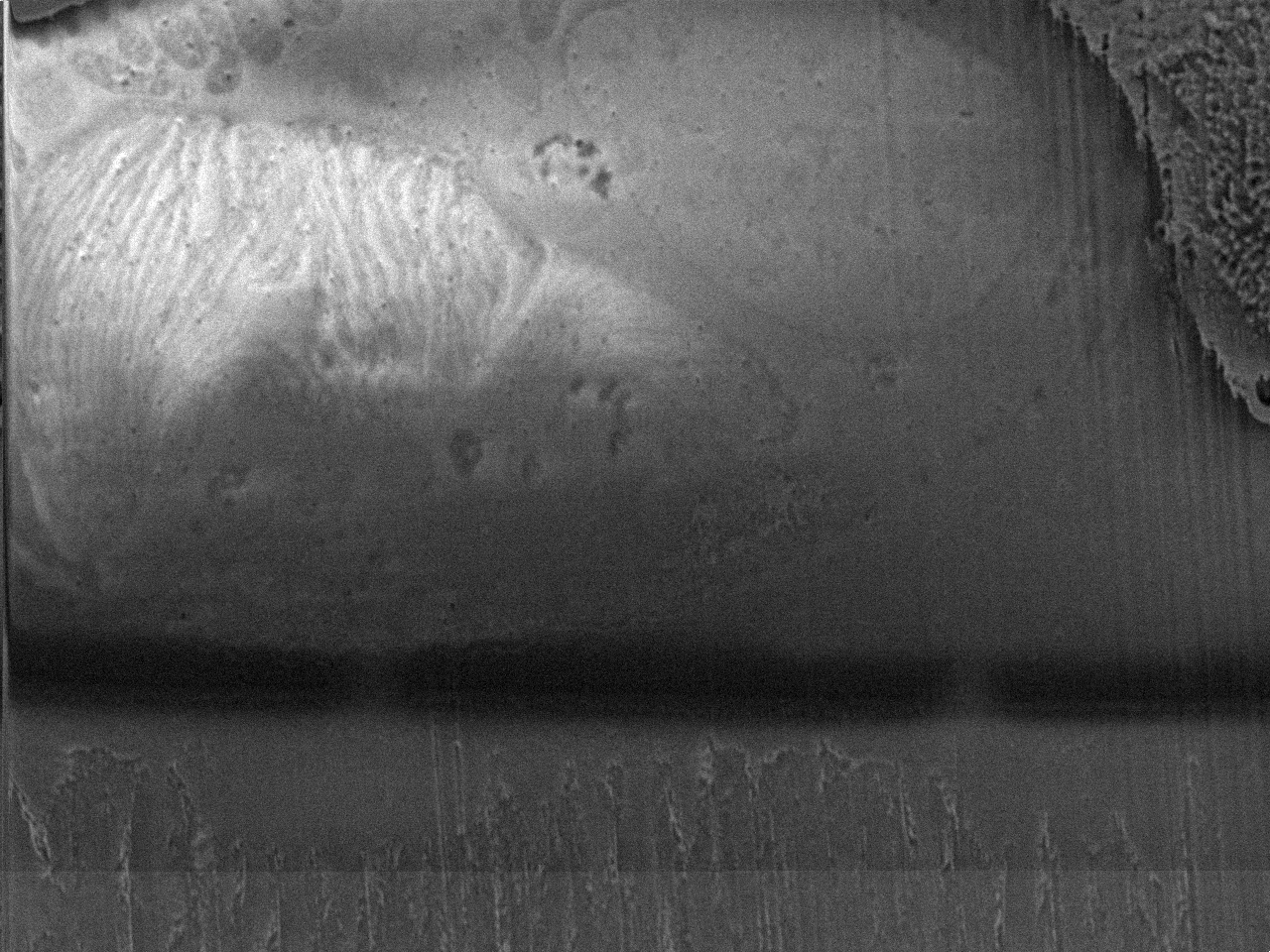}
        \end{minipage}
        \begin{minipage}{\colWidth\textwidth}
            \centering
            \includegraphics[width=\textwidth]{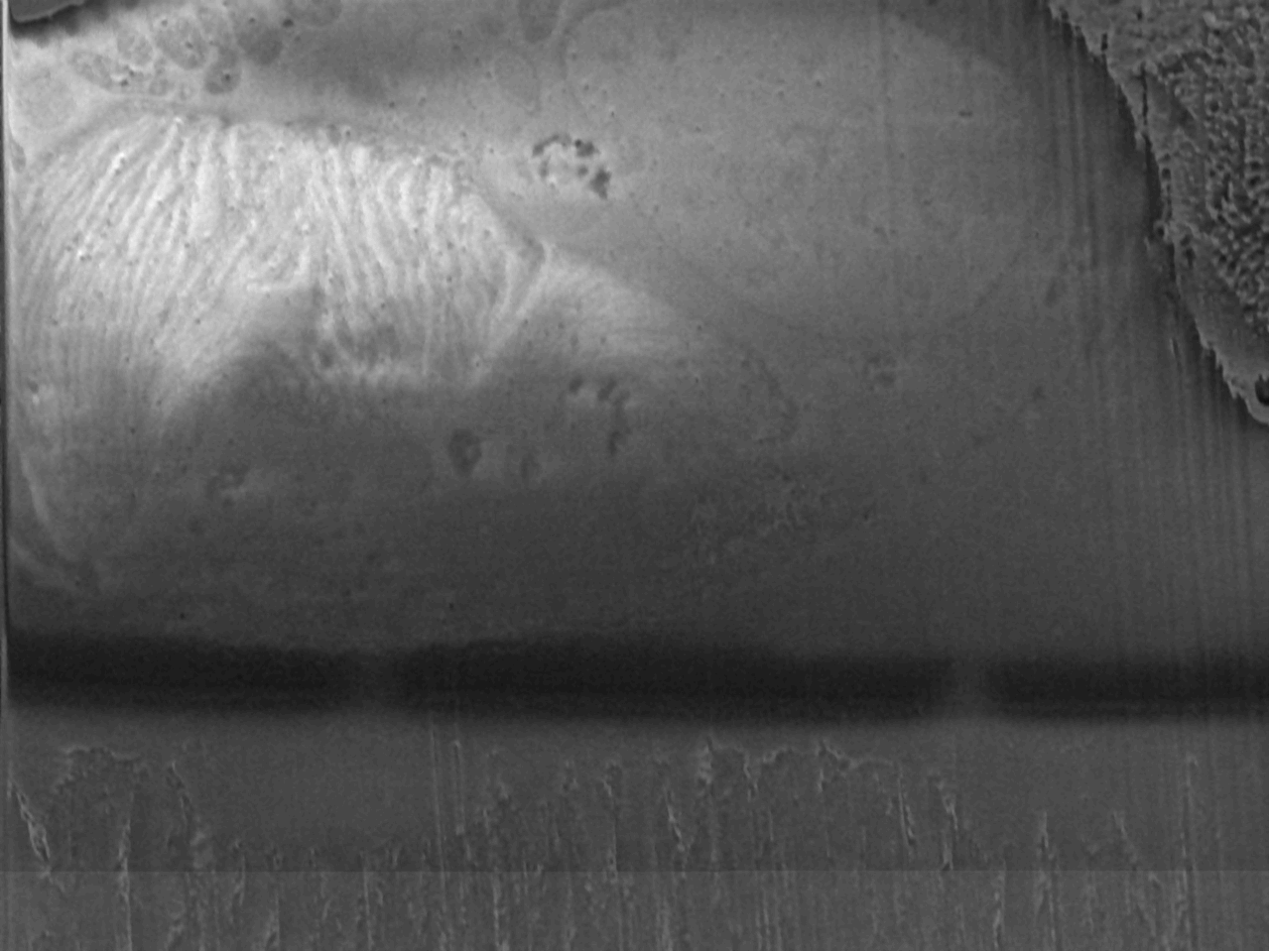}
        \end{minipage}
       
    \end{minipage}

    \vspace{\vertSpacing}
    
    \begin{minipage}{\textwidth}
        \centering
 
        \begin{minipage}{\firstColWidth\textwidth}
            \begin{center}
                50\%
            \end{center}
        \end{minipage}
        \begin{minipage}{\colWidth\textwidth}
            \centering
            \includegraphics[width=\textwidth]{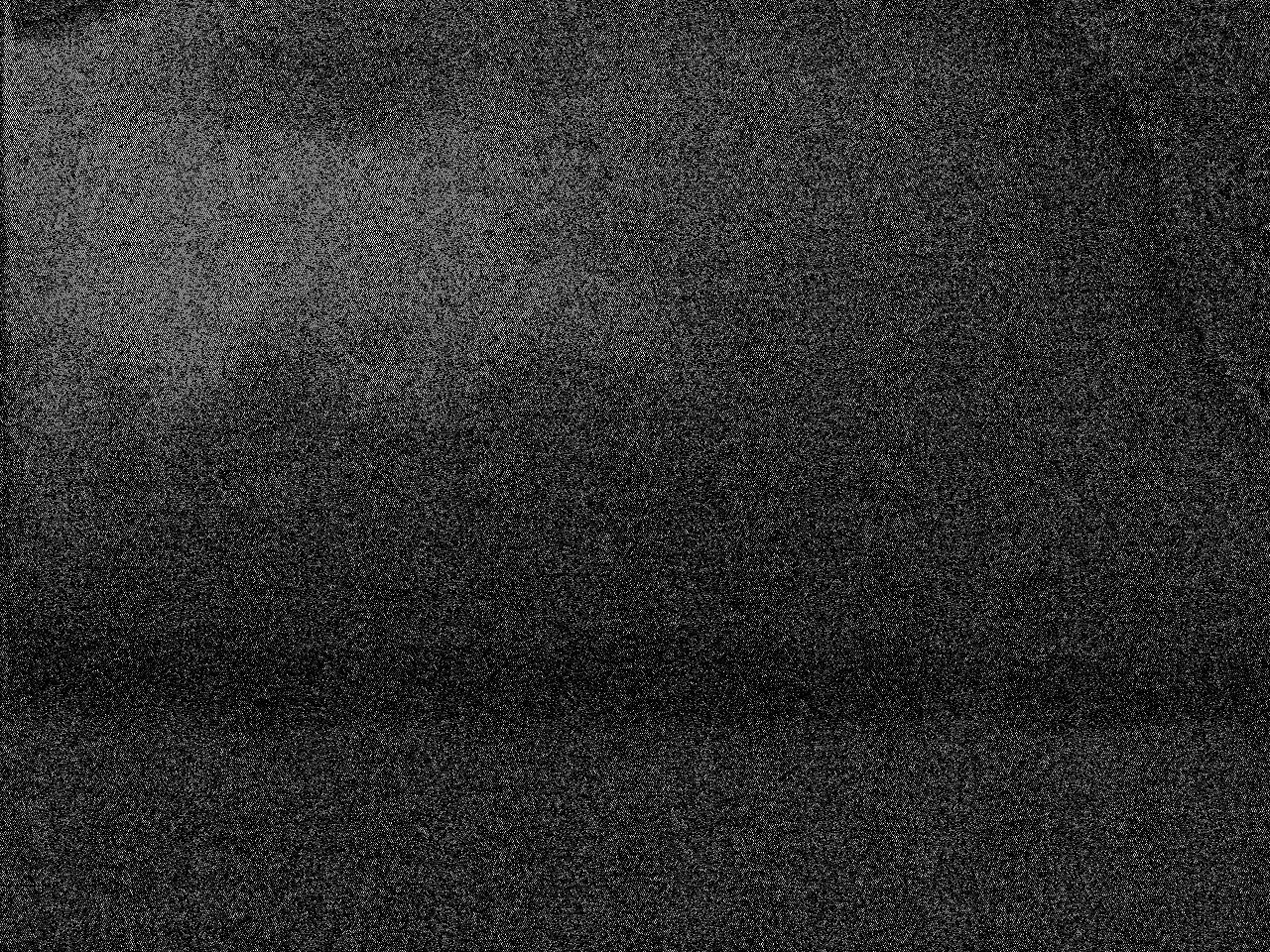}
        \end{minipage}        
        \begin{minipage}{\colWidth\textwidth}
            \centering
            \includegraphics[width=\textwidth]{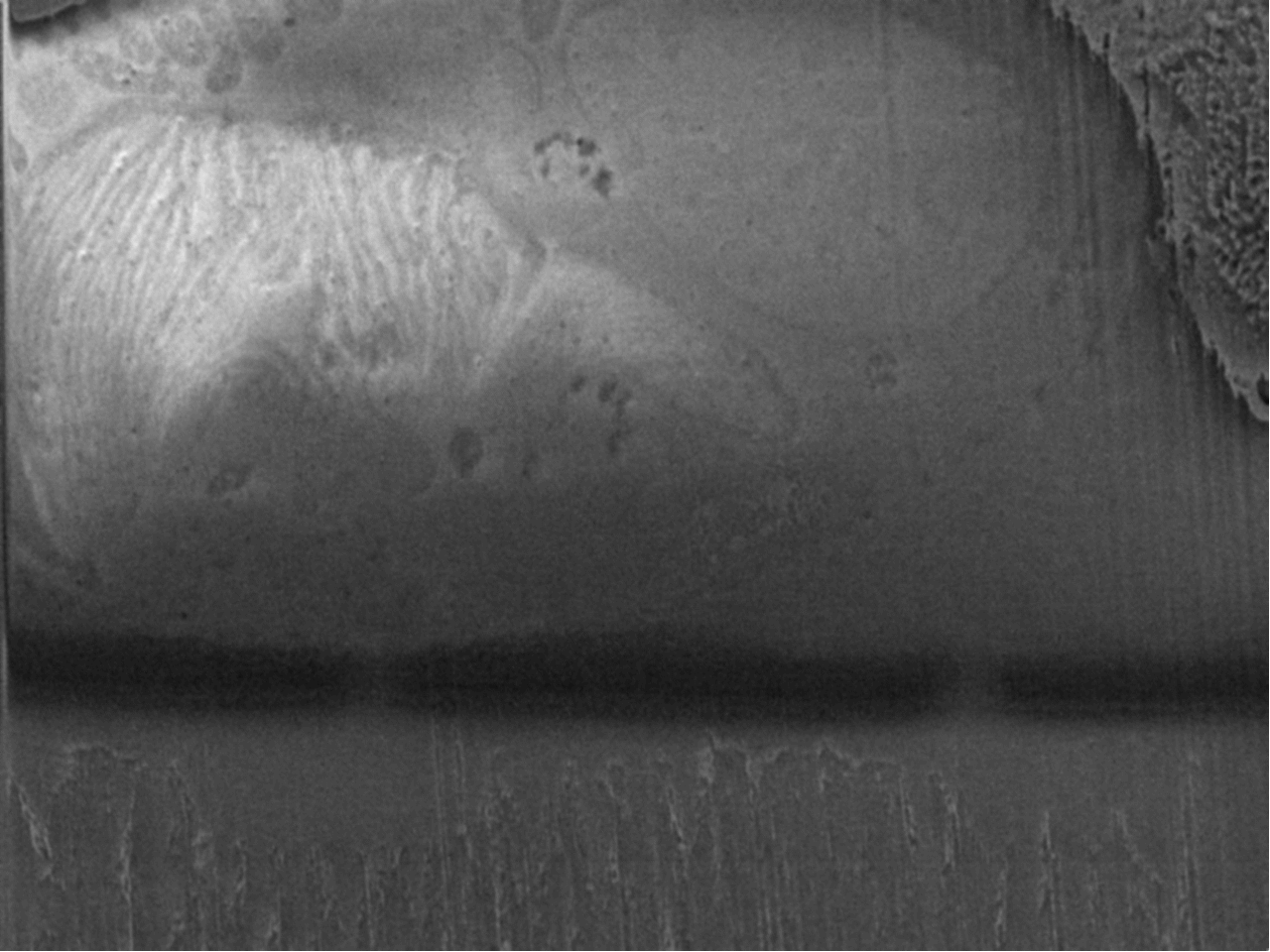}
        \end{minipage}
        \begin{minipage}{\colWidth\textwidth}
            \centering
            \includegraphics[width=\textwidth]{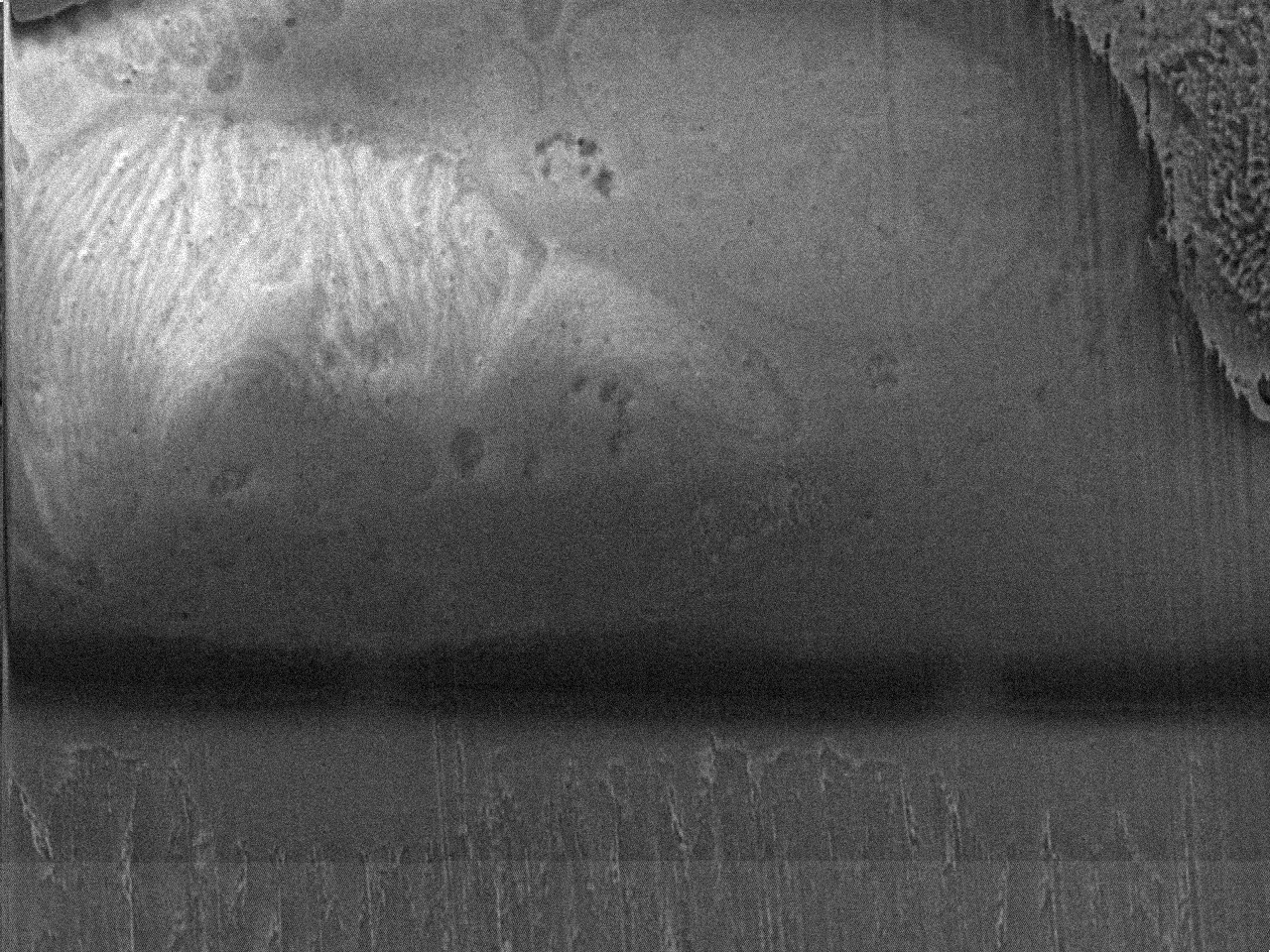}
        \end{minipage}
        \begin{minipage}{\colWidth\textwidth}
            \centering
            \includegraphics[width=\textwidth]{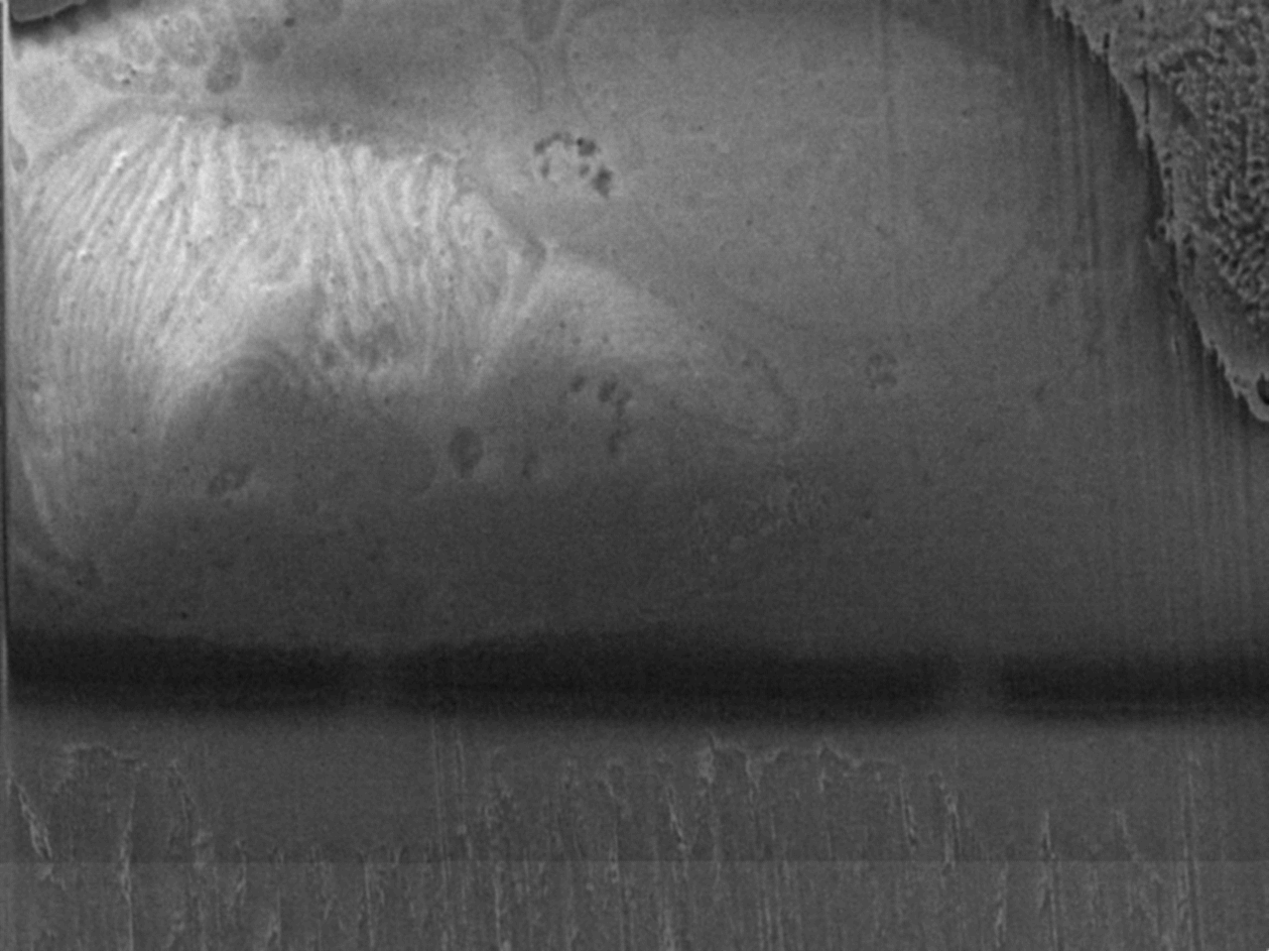}
        \end{minipage}
       
    \end{minipage}

    \vspace{\vertSpacing}
    
    \begin{minipage}{\textwidth}
        \centering
 
        \begin{minipage}{\firstColWidth\textwidth}
            \begin{center}
                25\%
            \end{center}
        \end{minipage}
        \begin{minipage}{\colWidth\textwidth}
            \centering
            \includegraphics[width=\textwidth]{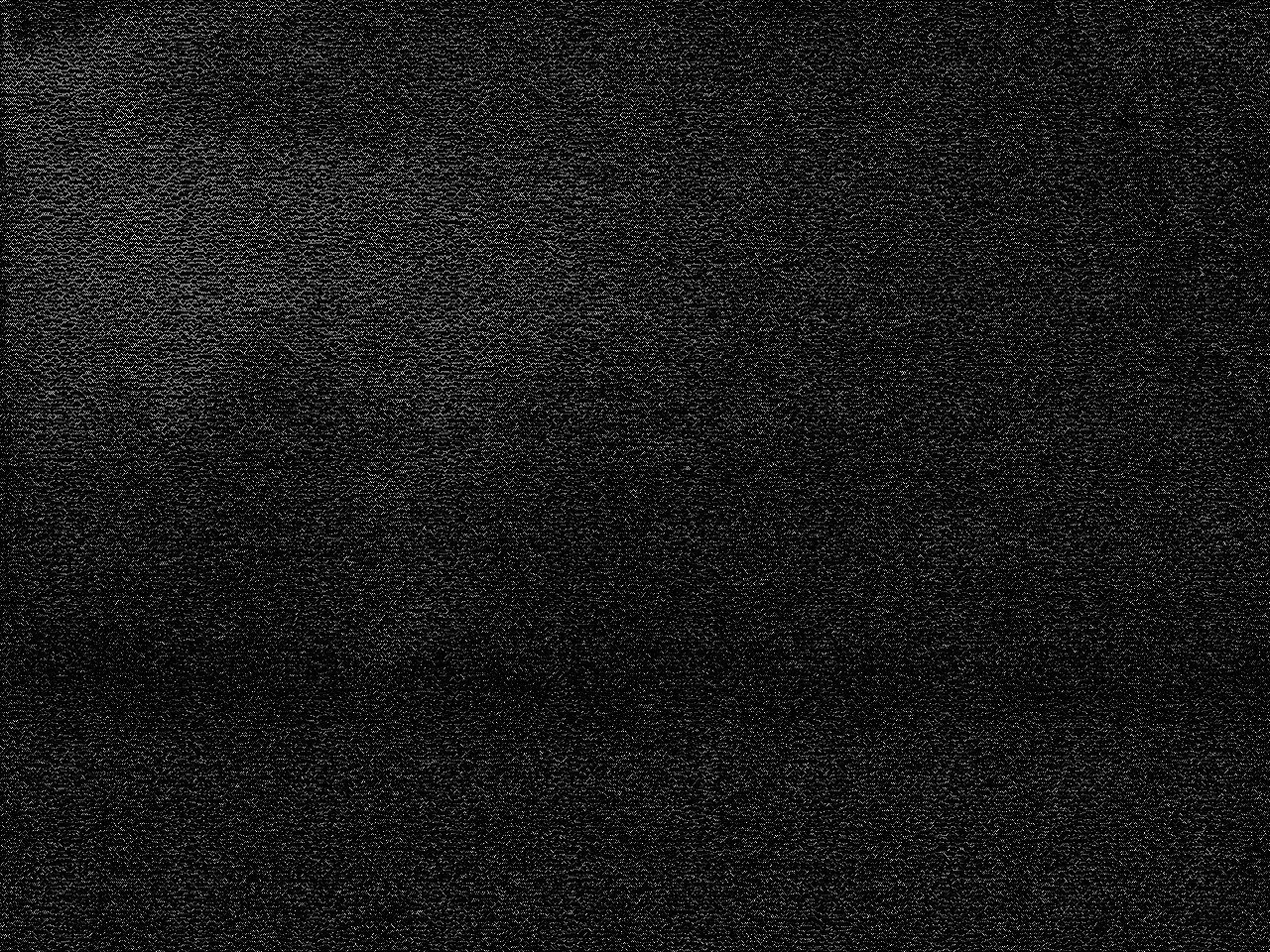}
        \end{minipage}        
        \begin{minipage}{\colWidth\textwidth}
            \centering
            \includegraphics[width=\textwidth]{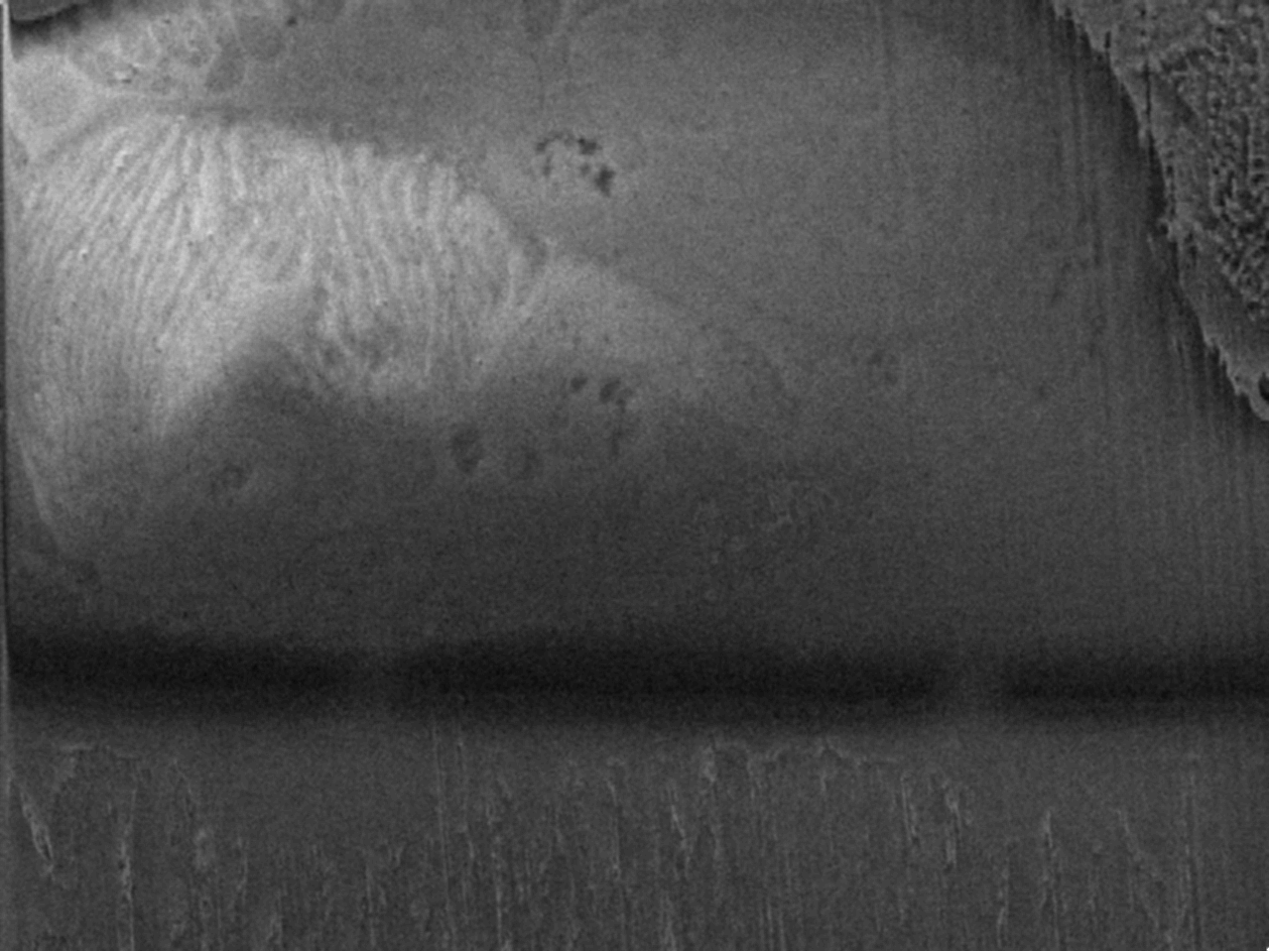}
        \end{minipage}
        \begin{minipage}{\colWidth\textwidth}
            \centering
            \includegraphics[width=\textwidth]{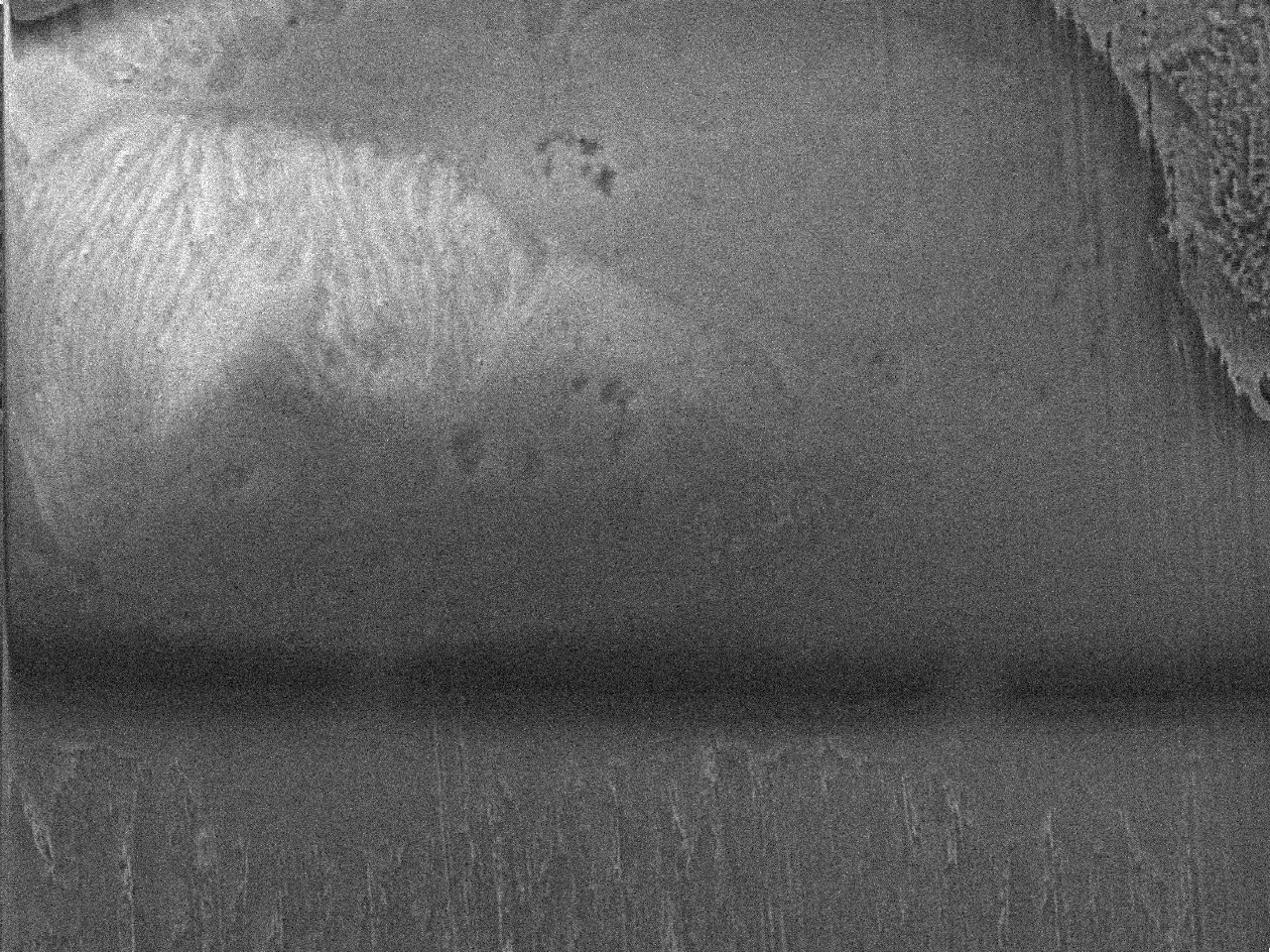}
        \end{minipage}
        \begin{minipage}{\colWidth\textwidth}
            \centering
            \includegraphics[width=\textwidth]{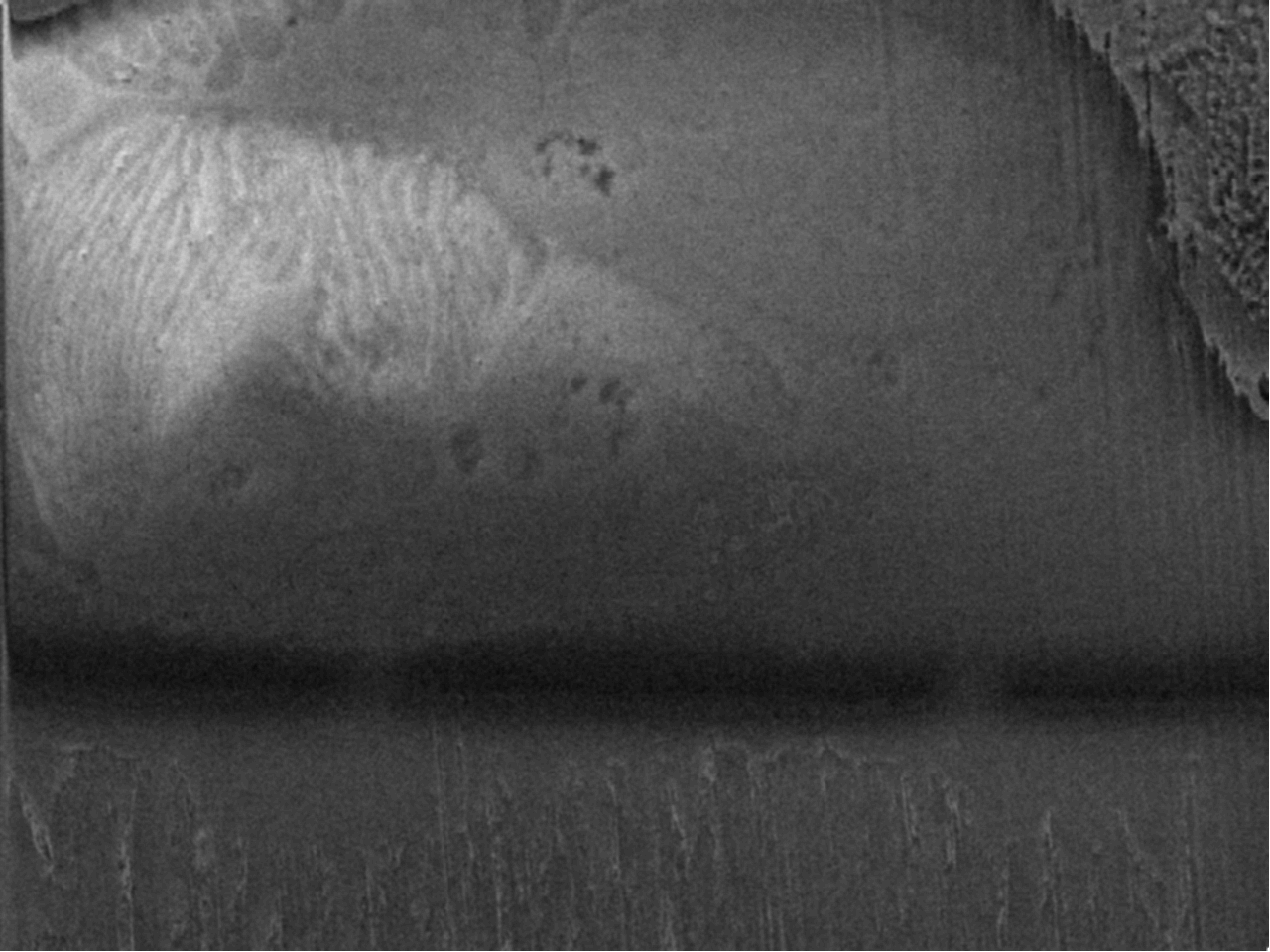}
        \end{minipage}
       
    \end{minipage}

    \vspace{\vertSpacing}
    
    \begin{minipage}{\textwidth}
        \centering
 
        \begin{minipage}{\firstColWidth\textwidth}
            \begin{center}
                20\%
            \end{center}
        \end{minipage}
        \begin{minipage}{\colWidth\textwidth}
            \centering
            \includegraphics[width=\textwidth]{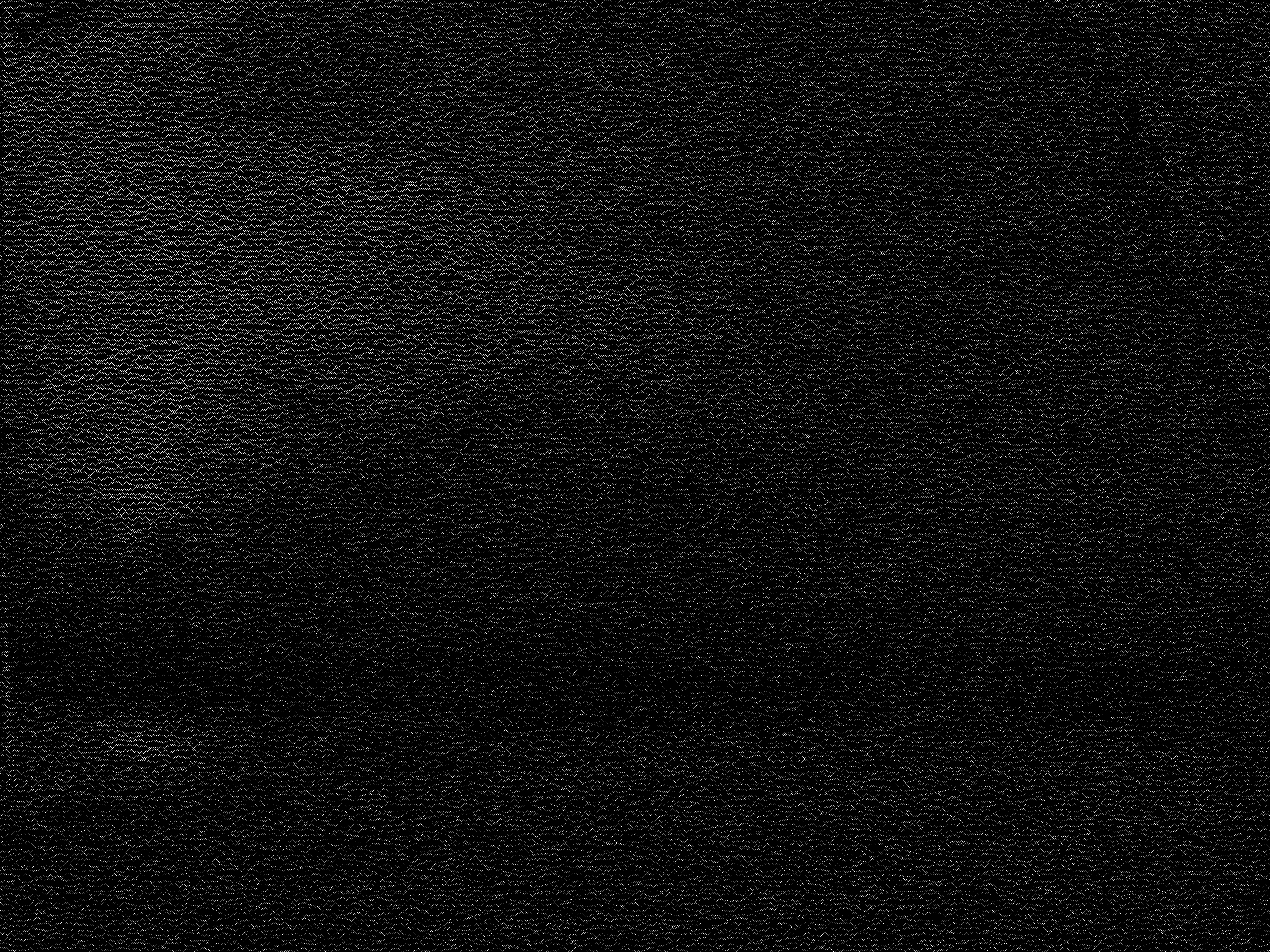}
        \end{minipage}        
        \begin{minipage}{\colWidth\textwidth}
            \centering
            \includegraphics[width=\textwidth]{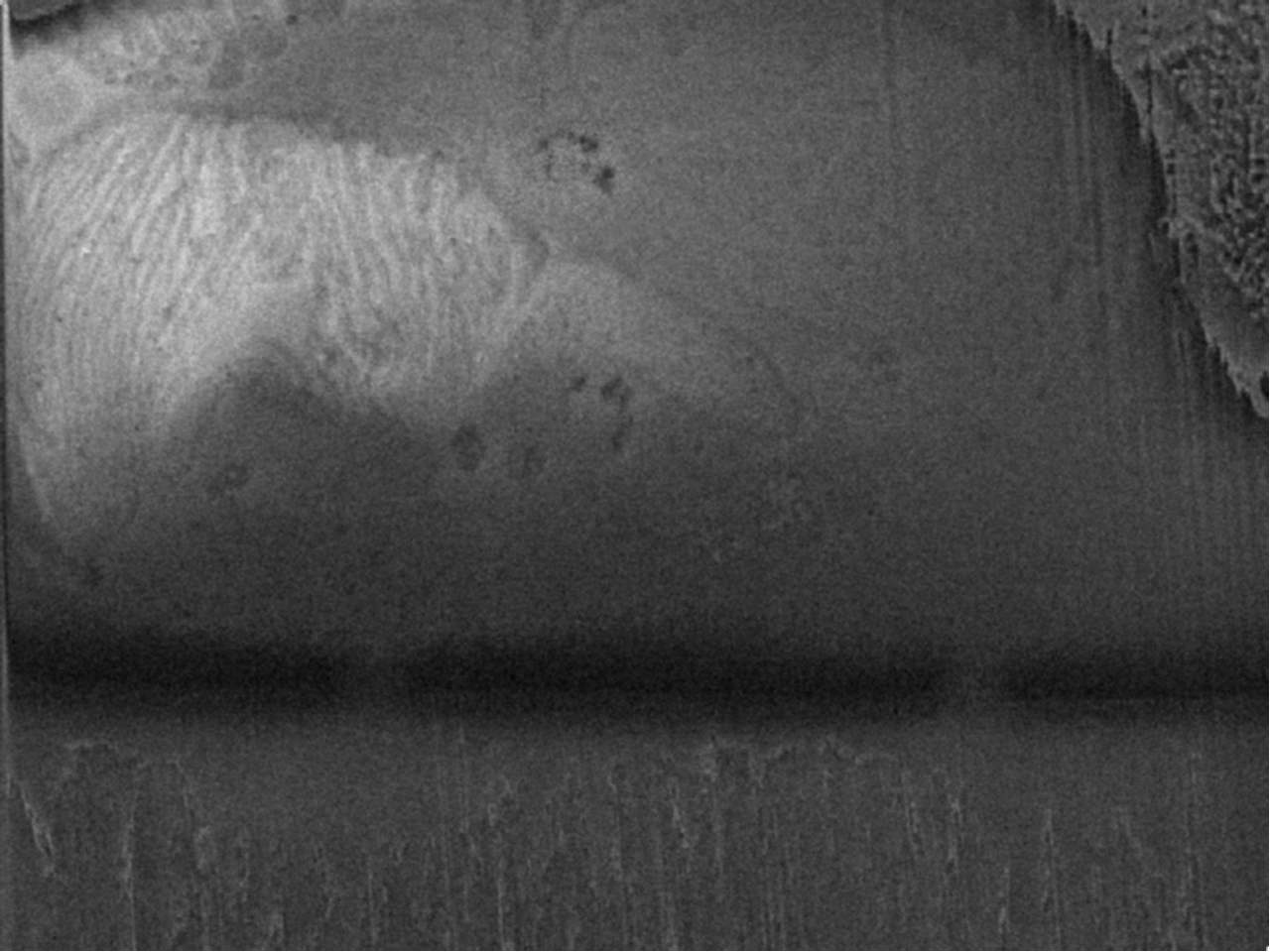}
        \end{minipage}
        \begin{minipage}{\colWidth\textwidth}
            \centering
            \includegraphics[width=\textwidth]{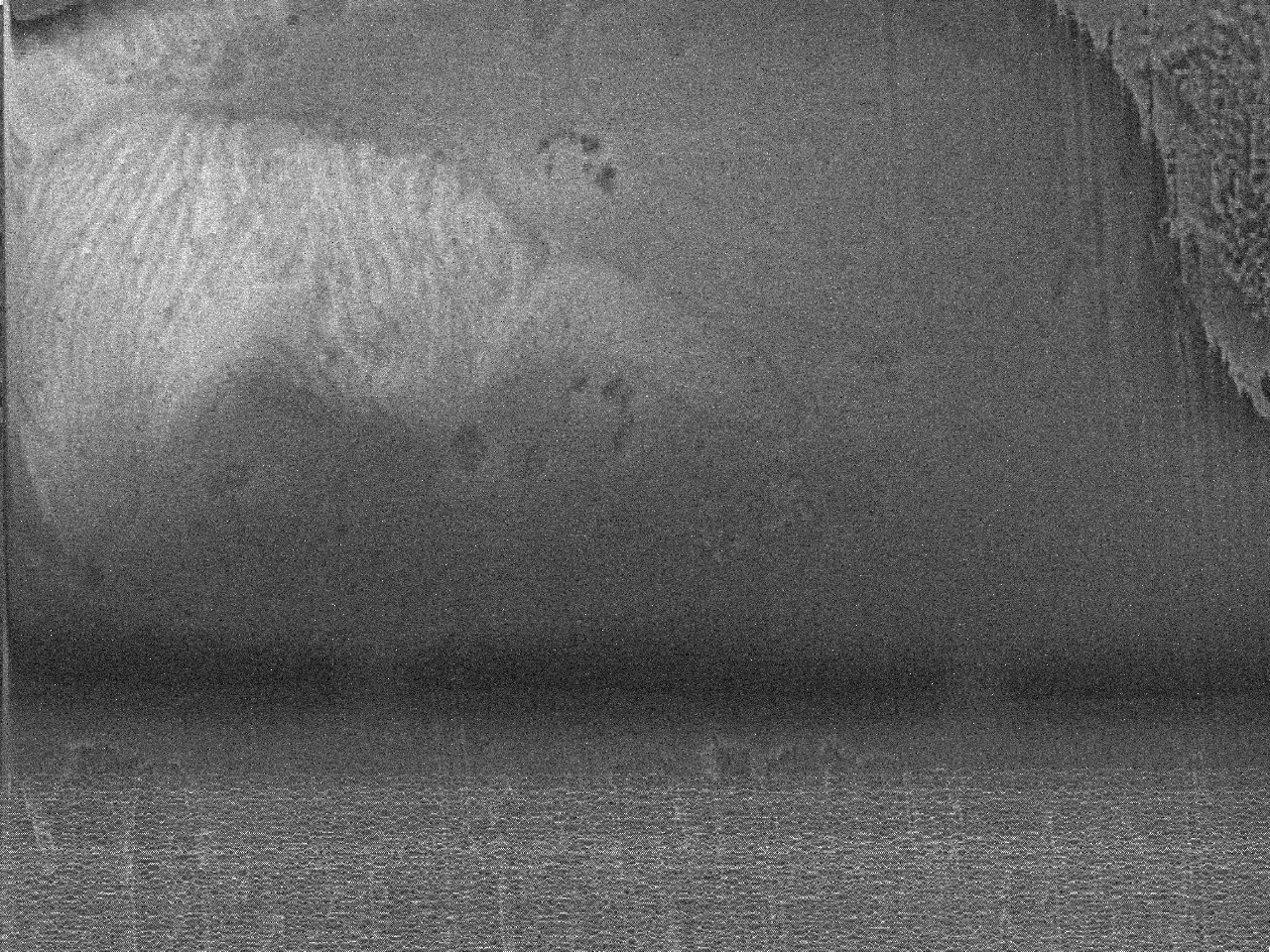}
        \end{minipage}
        \begin{minipage}{\colWidth\textwidth}
            \centering
            \includegraphics[width=\textwidth]{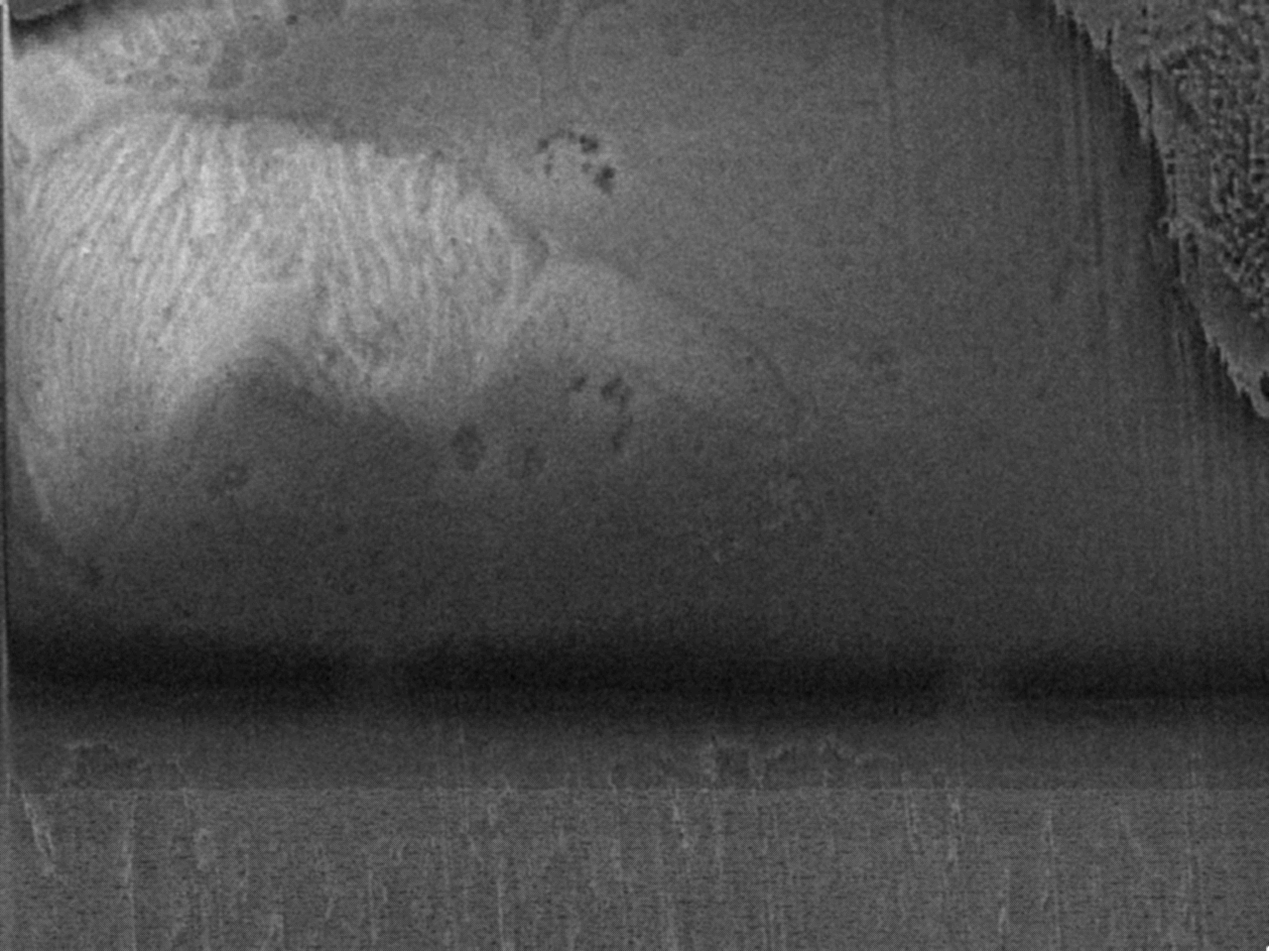}
        \end{minipage}
       
    \end{minipage}

    \vspace{\vertSpacing}
    
    \begin{minipage}{\textwidth}
        \centering
 
        \begin{minipage}{\firstColWidth\textwidth}
            \begin{center}
                15\%
            \end{center}
        \end{minipage}
        \begin{minipage}{\colWidth\textwidth}
            \centering
            \includegraphics[width=\textwidth]{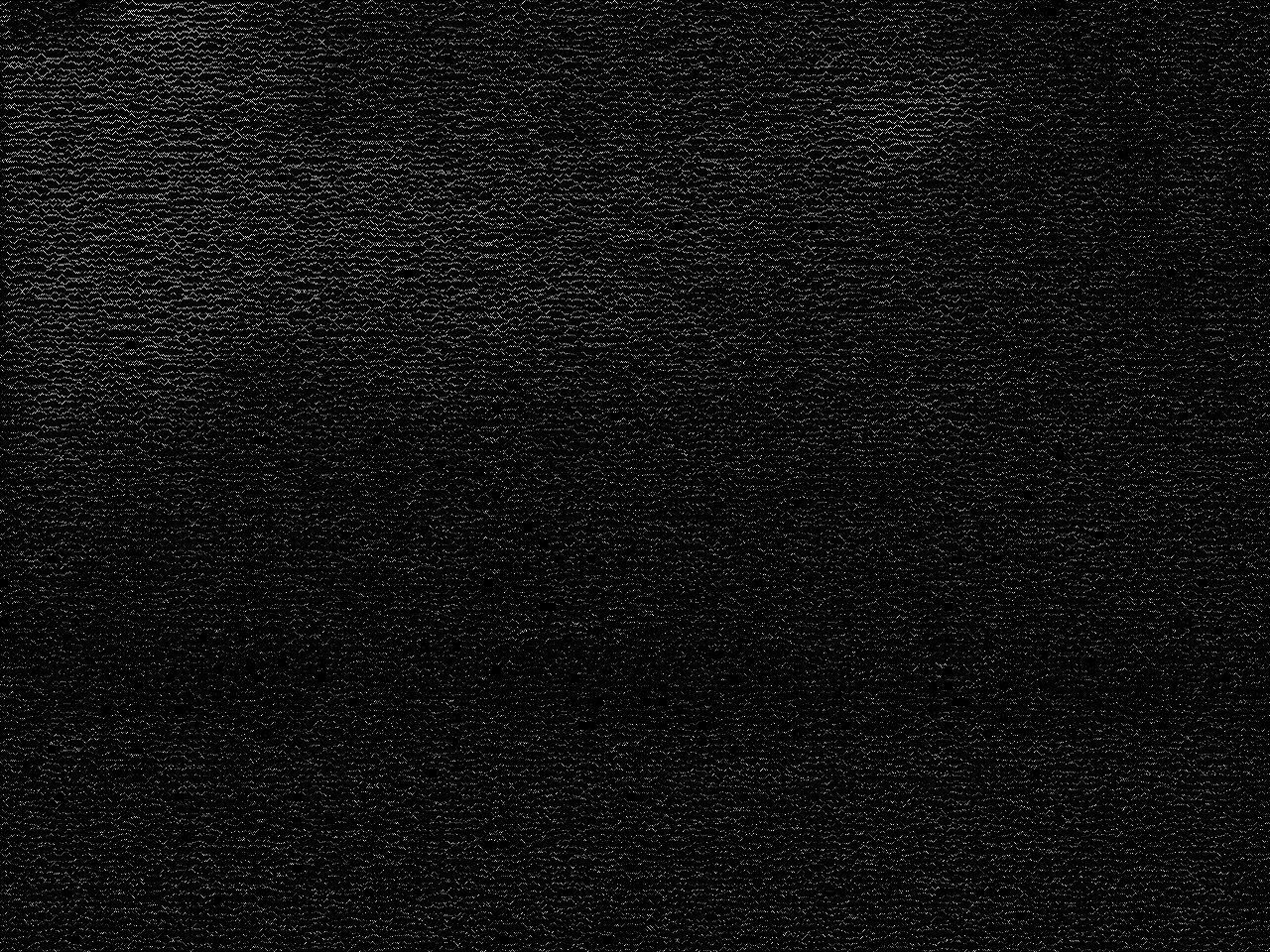}
        \end{minipage}        
        \begin{minipage}{\colWidth\textwidth}
            \centering
            \includegraphics[width=\textwidth]{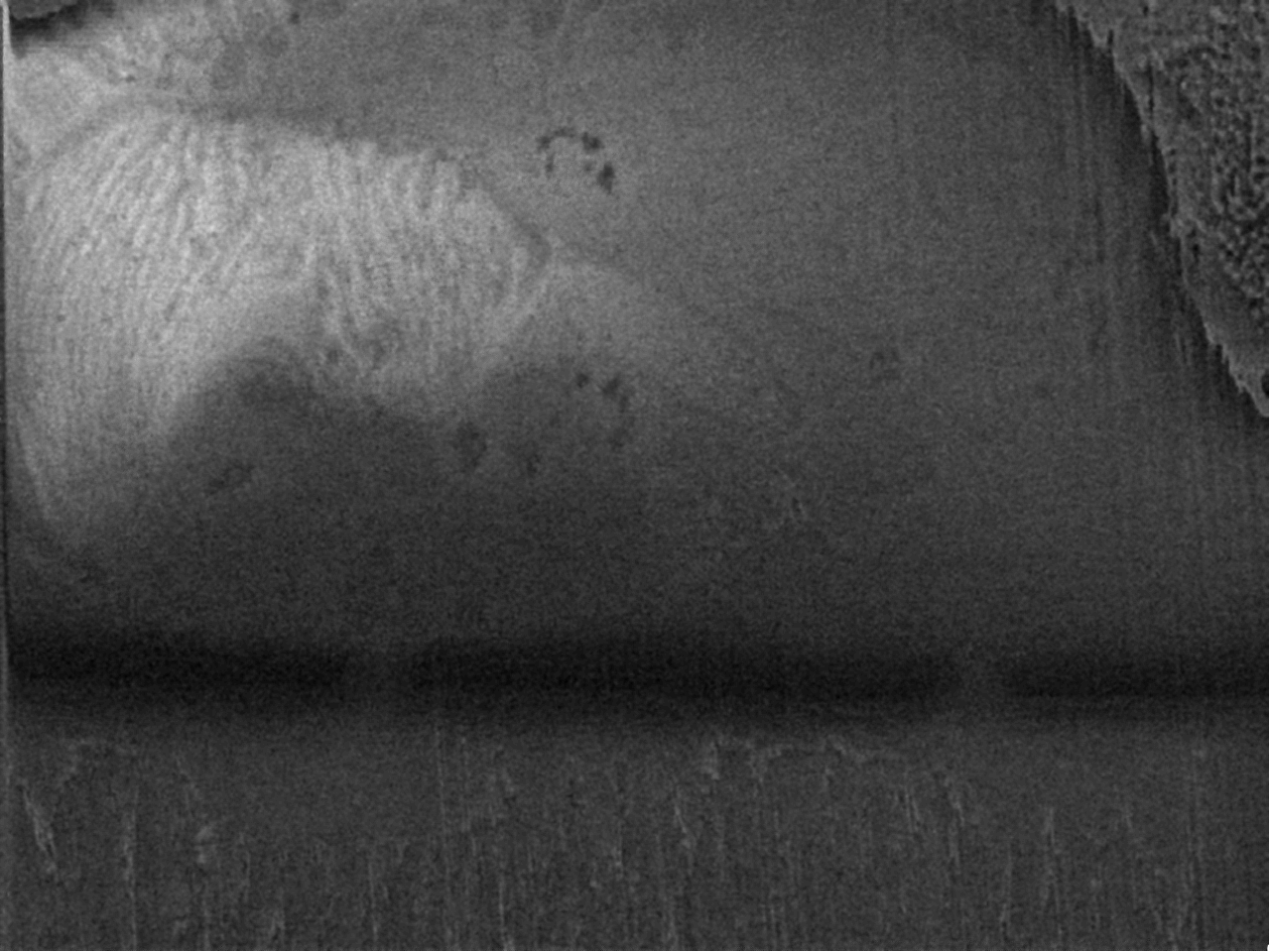}
        \end{minipage}
        \begin{minipage}{\colWidth\textwidth}
            \centering
            \includegraphics[width=\textwidth]{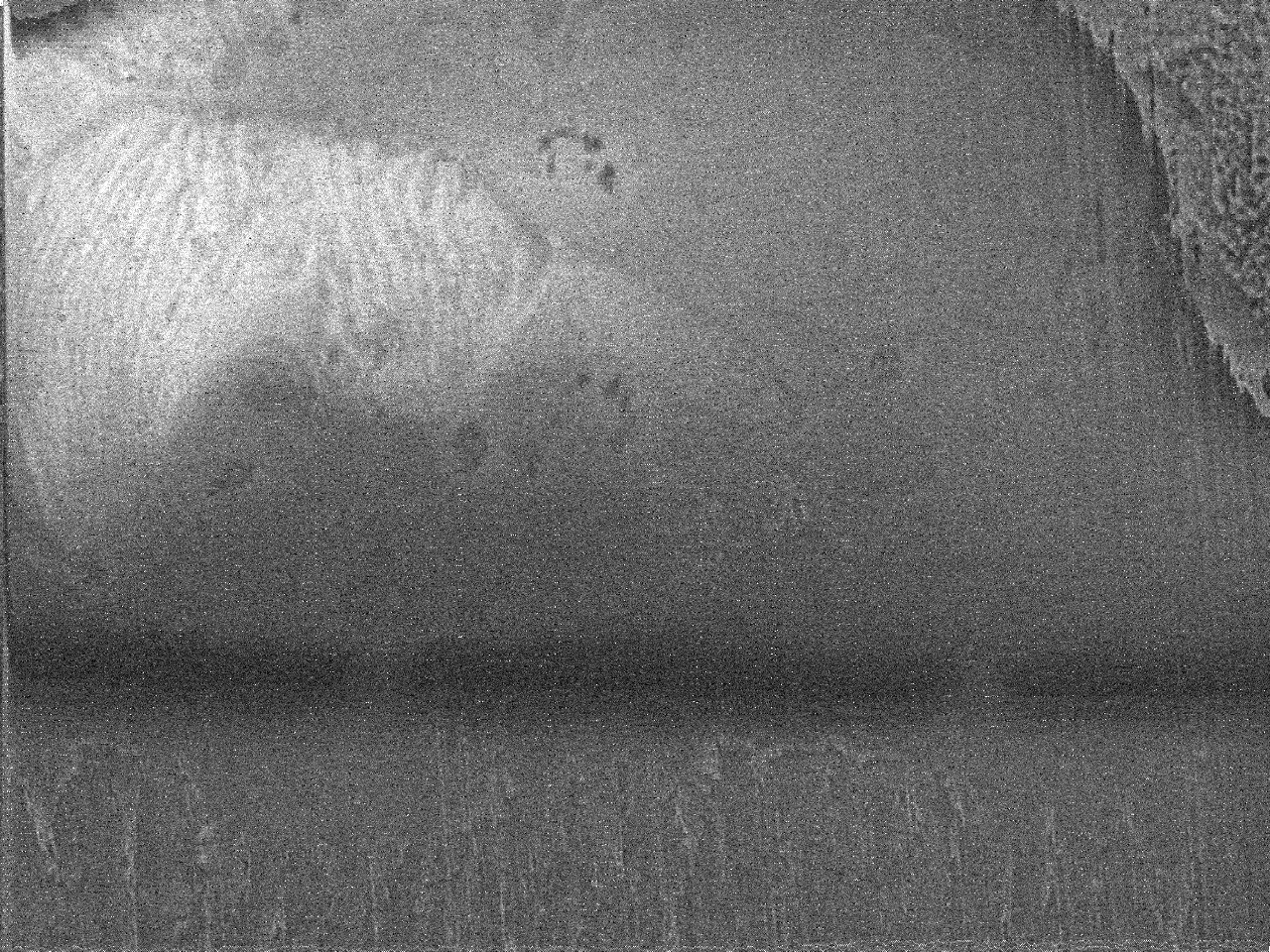}
        \end{minipage}
        \begin{minipage}{\colWidth\textwidth}
            \centering
            \includegraphics[width=\textwidth]{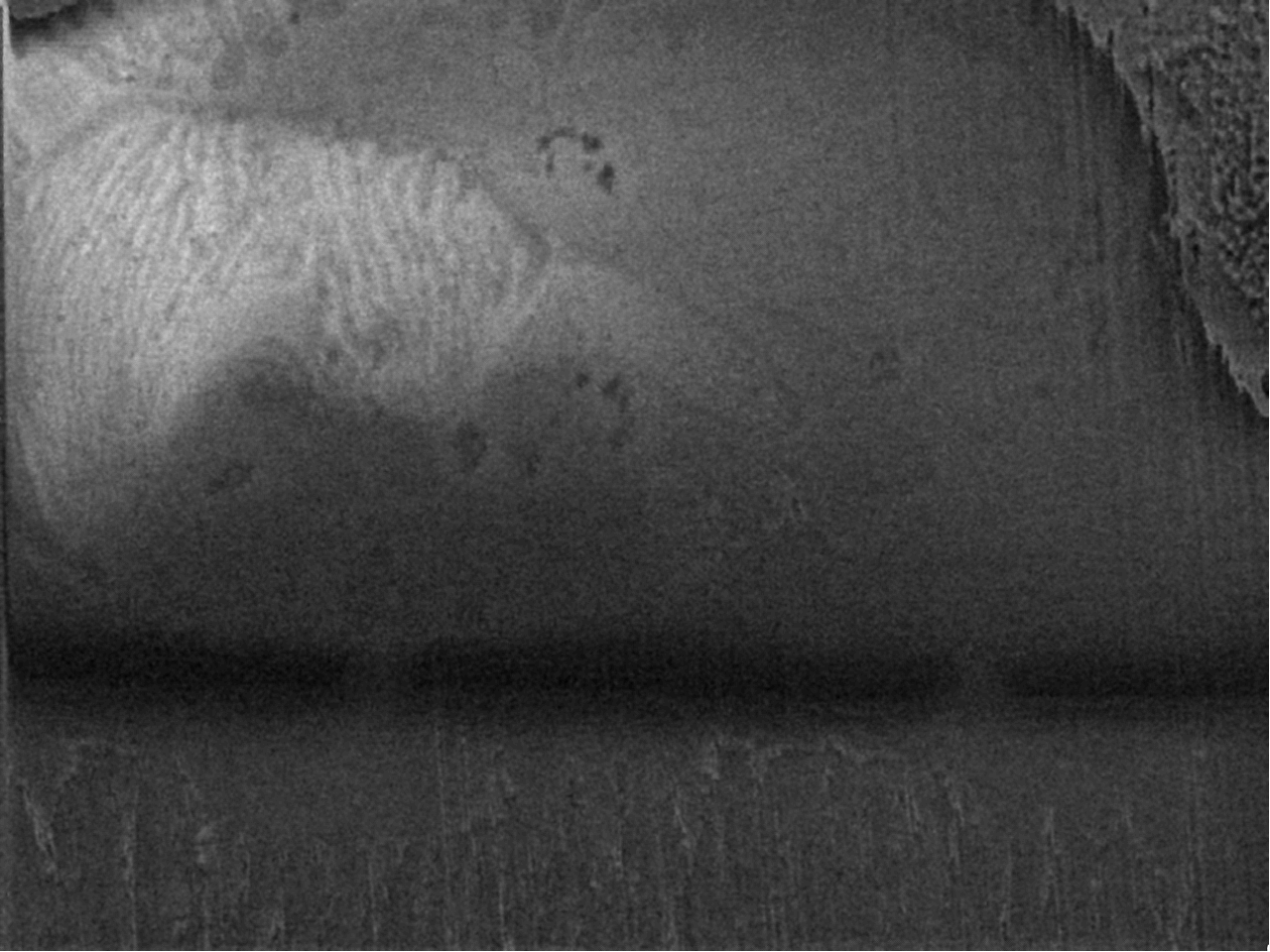}
        \end{minipage}
       
    \end{minipage}

    \vspace{\vertSpacing}
    
    \begin{minipage}{\textwidth}
        \centering
 
        \begin{minipage}{\firstColWidth\textwidth}
            \begin{center}
                10\%
            \end{center}
        \end{minipage}
        \begin{minipage}{\colWidth\textwidth}
            \centering
            \includegraphics[width=\textwidth]{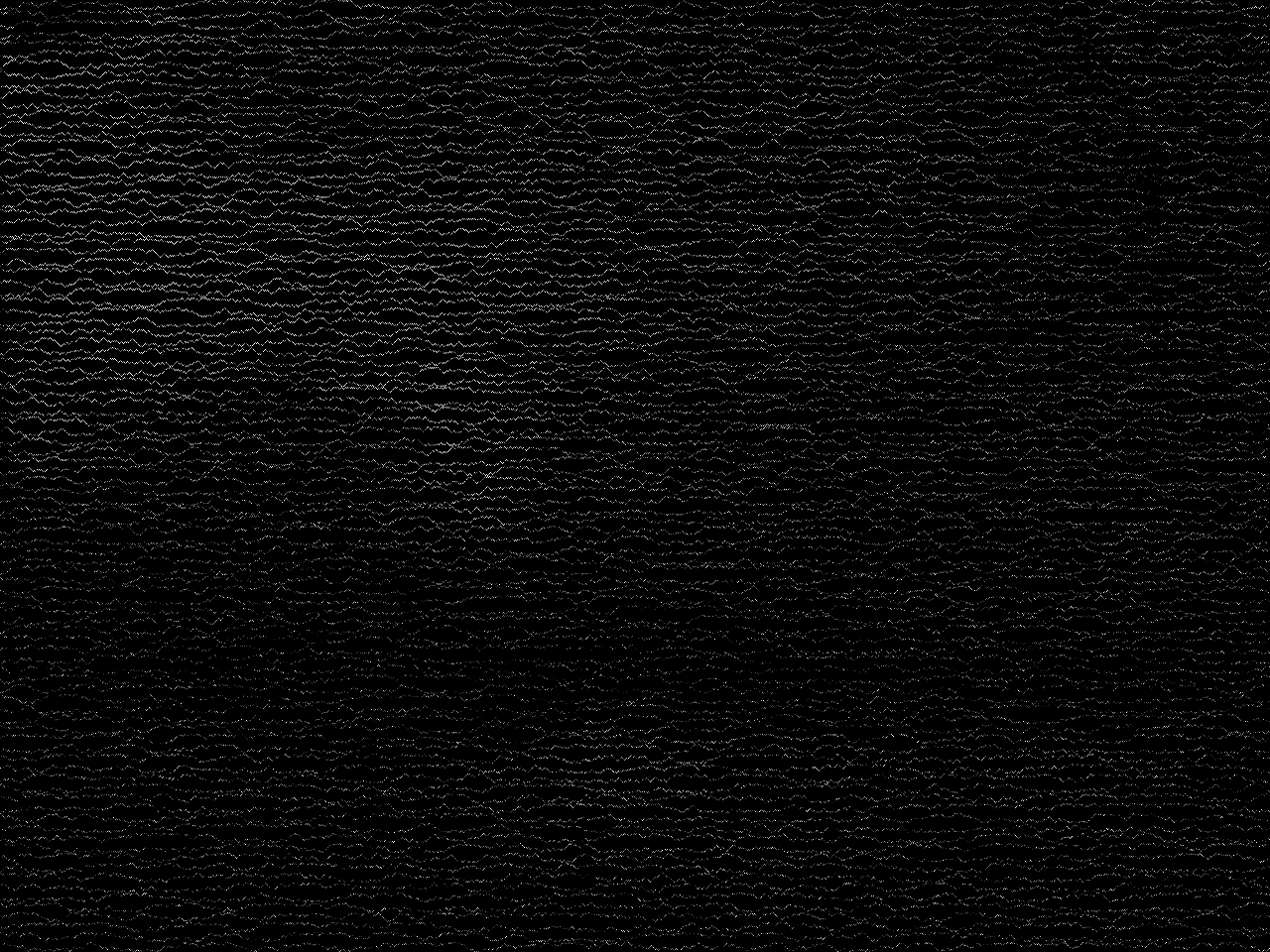}
        \end{minipage}        
        \begin{minipage}{\colWidth\textwidth}
            \centering
            \includegraphics[width=\textwidth]{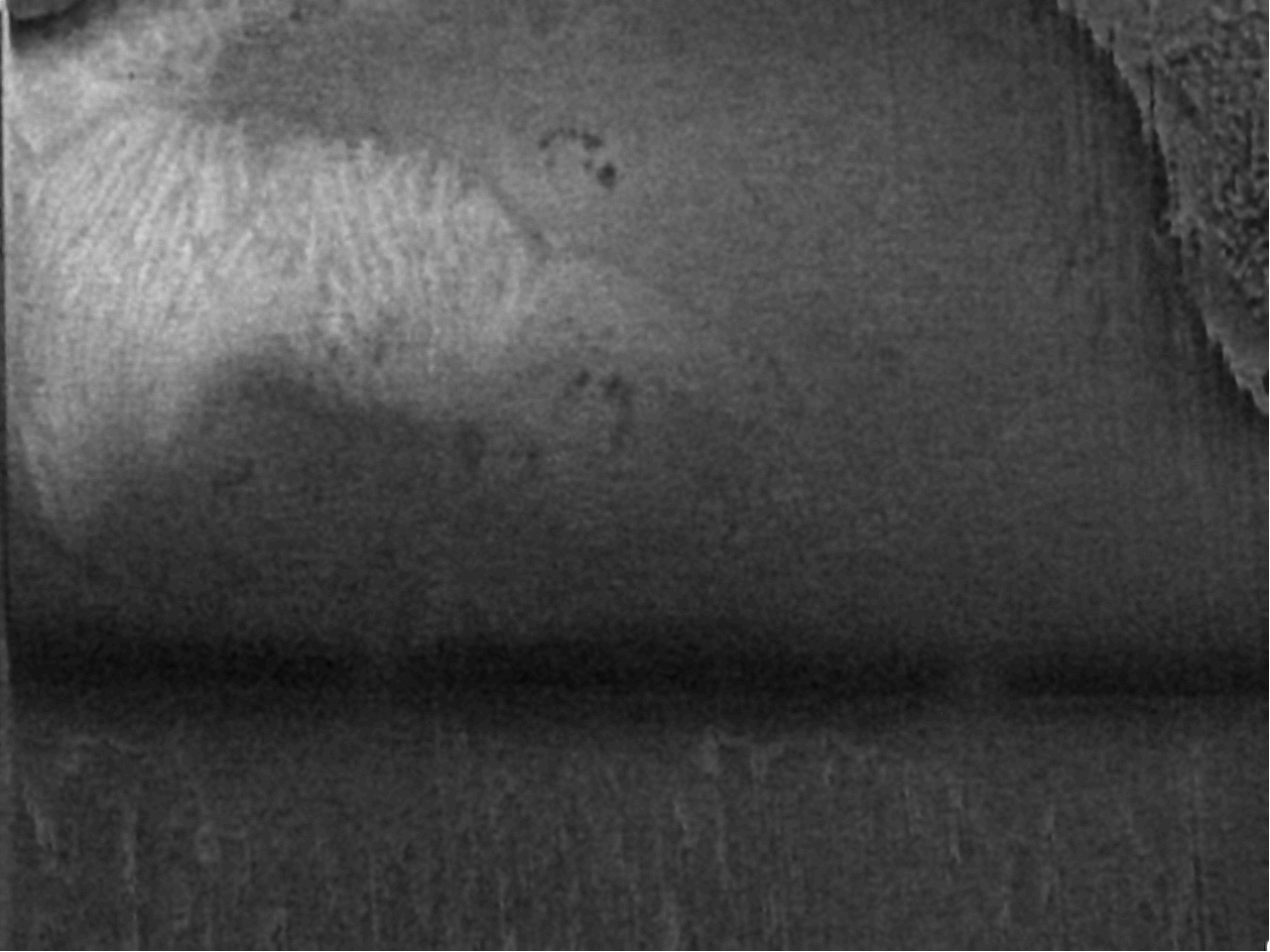}
        \end{minipage}
        \begin{minipage}{\colWidth\textwidth}
            \centering
            \includegraphics[width=\textwidth]{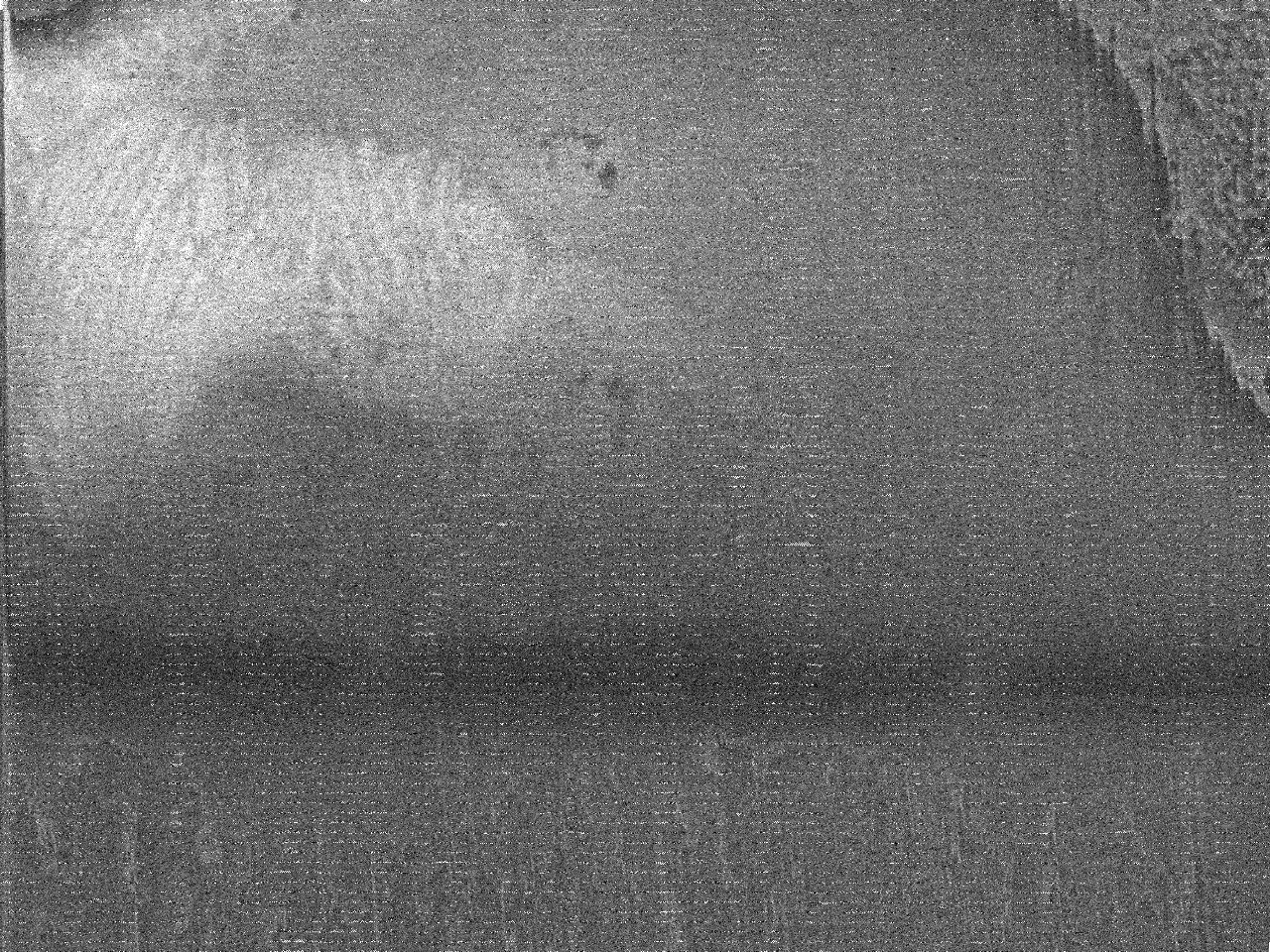}
        \end{minipage}
        \begin{minipage}{\colWidth\textwidth}
            \centering
            \includegraphics[width=\textwidth]{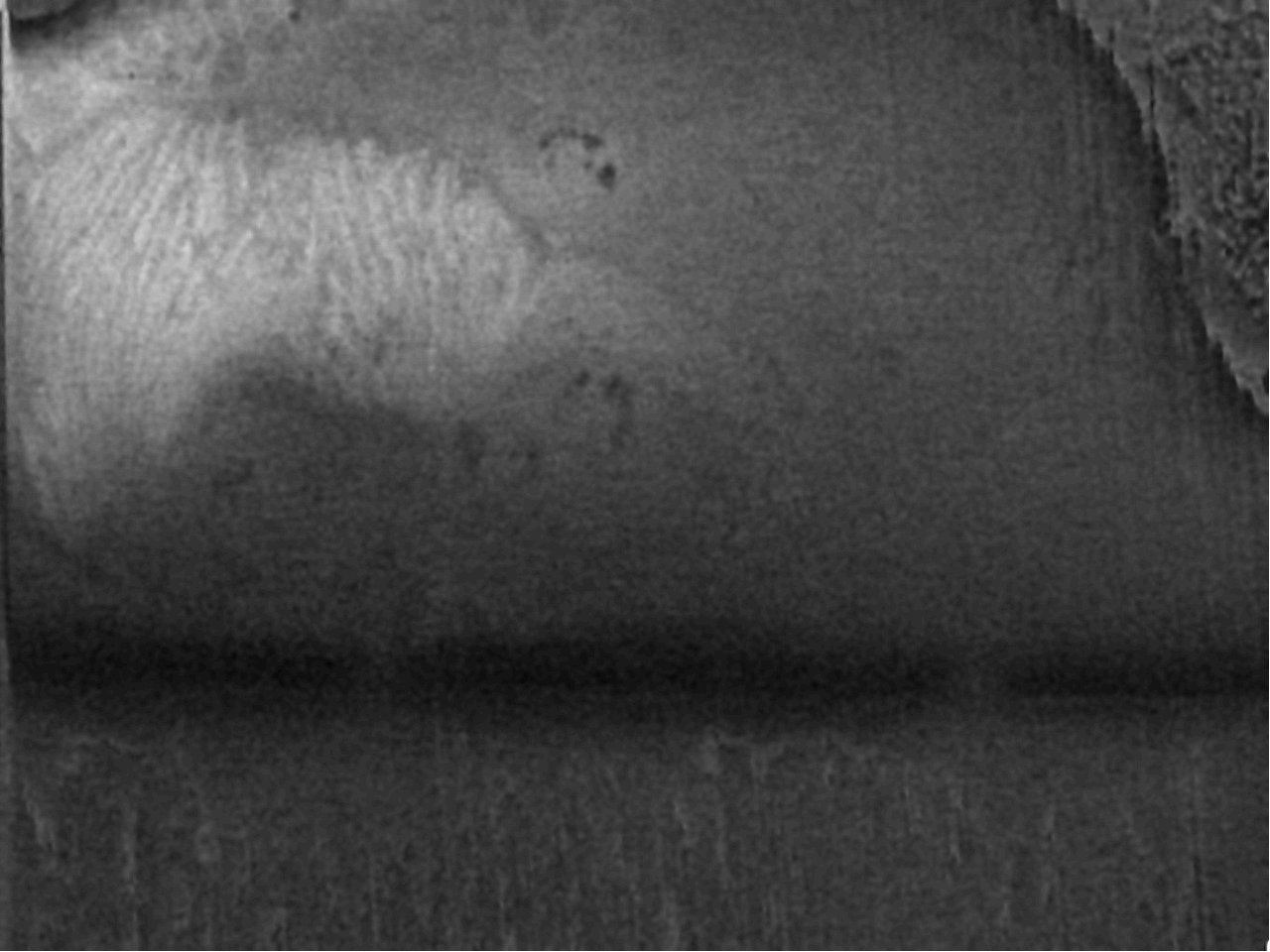}
        \end{minipage}
       
    \end{minipage}

    \vspace{\vertSpacing}
    
    \begin{minipage}{\textwidth}
        \centering
 
        \begin{minipage}{\firstColWidth\textwidth}
            \begin{center}
                5\%
            \end{center}
        \end{minipage}
        \begin{minipage}{\colWidth\textwidth}
            \centering
            \includegraphics[width=\textwidth]{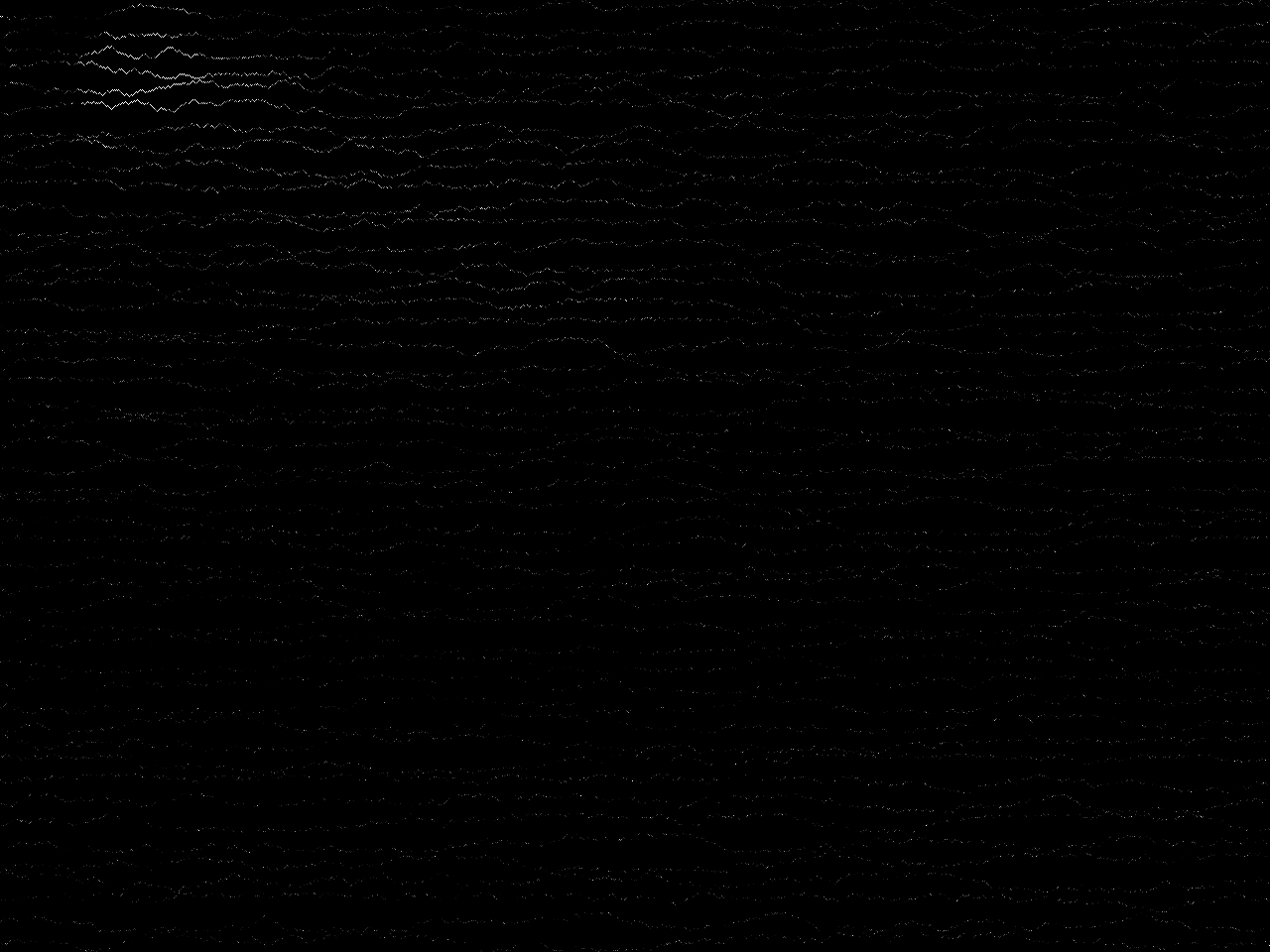}
        \end{minipage}        
        \begin{minipage}{\colWidth\textwidth}
            \centering
            \includegraphics[width=\textwidth]{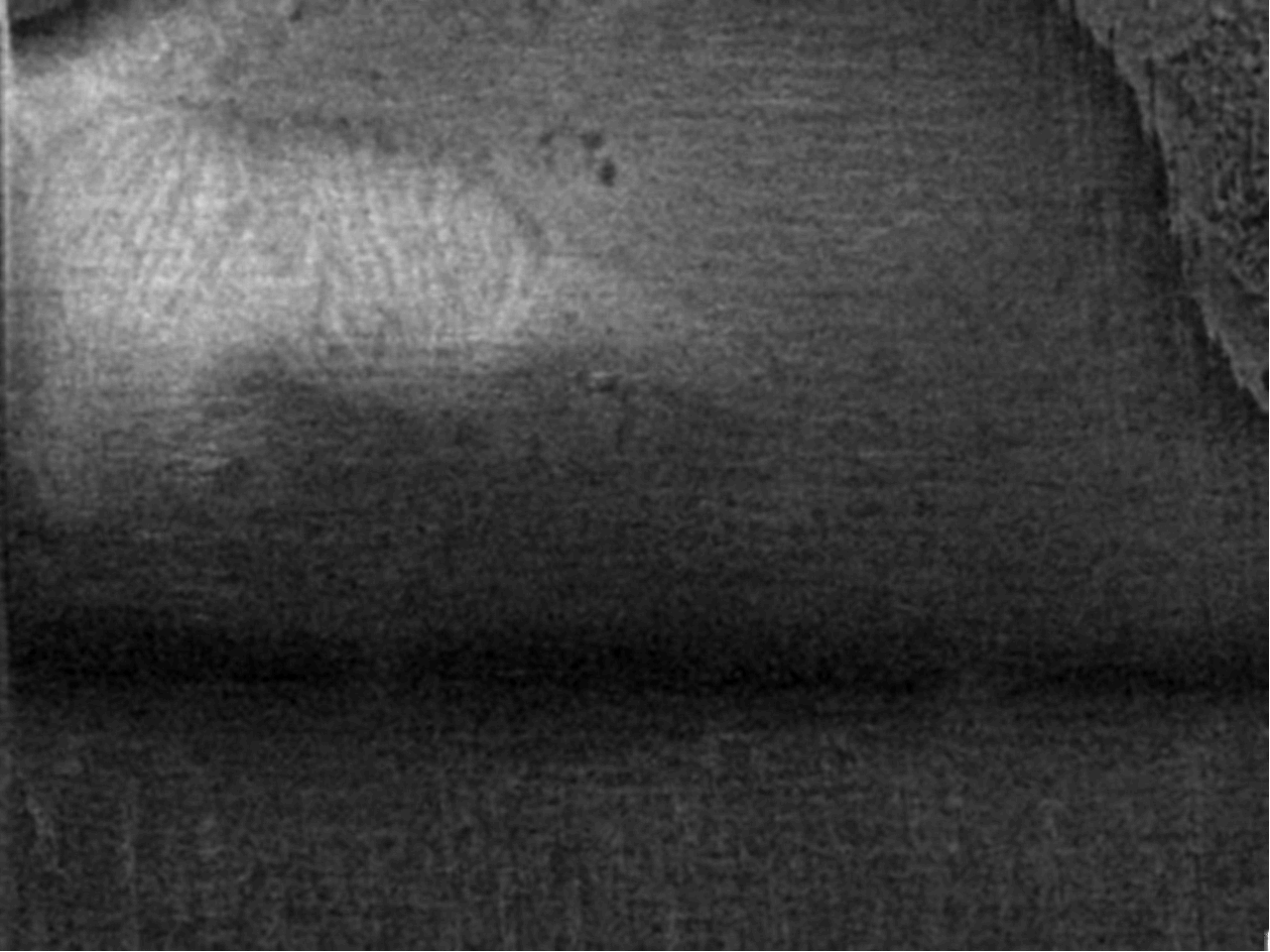}
        \end{minipage}
        \begin{minipage}{\colWidth\textwidth}
            \centering
            \includegraphics[width=\textwidth]{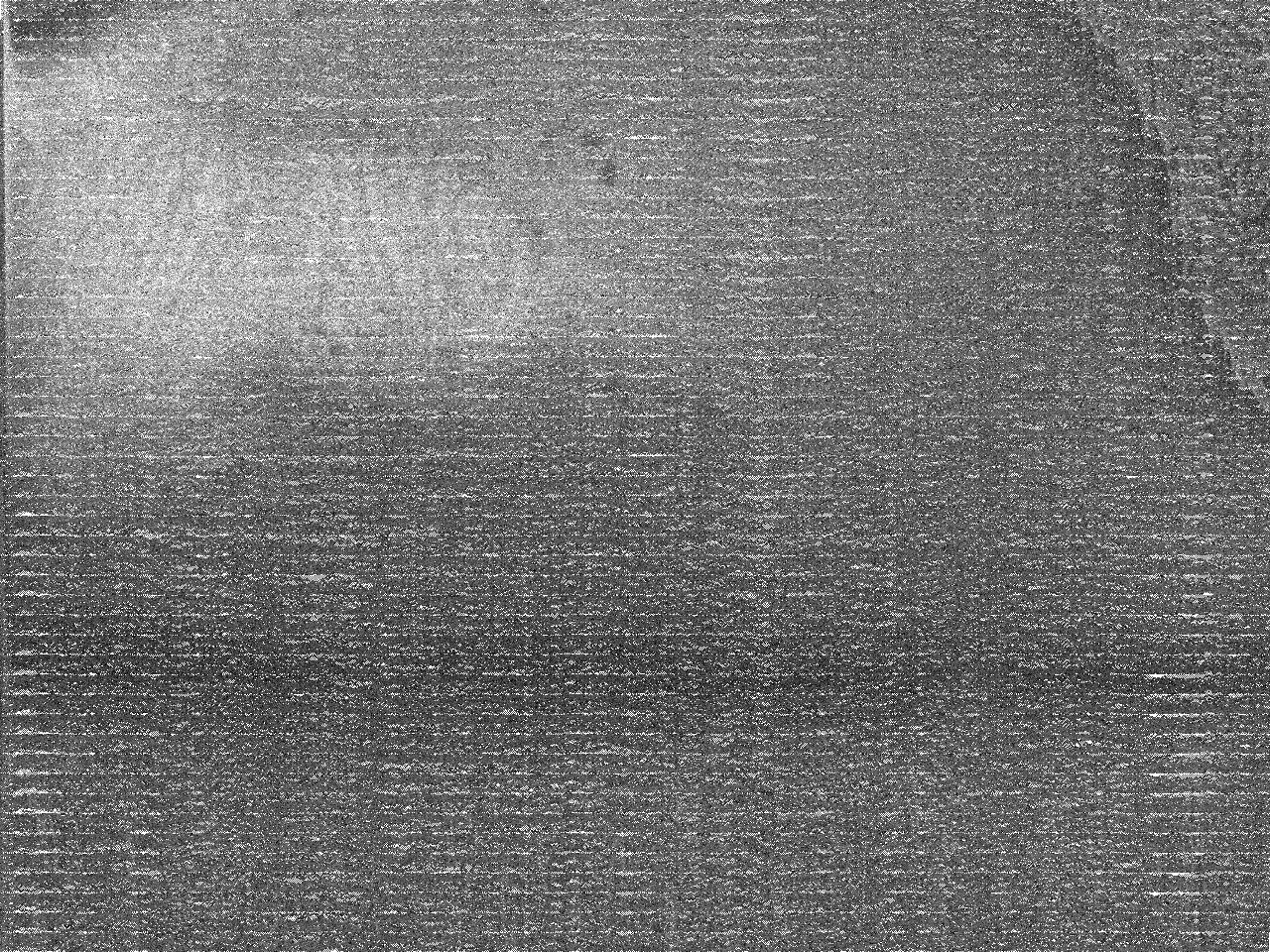}
        \end{minipage}
        \begin{minipage}{\colWidth\textwidth}
            \centering
            \includegraphics[width=\textwidth]{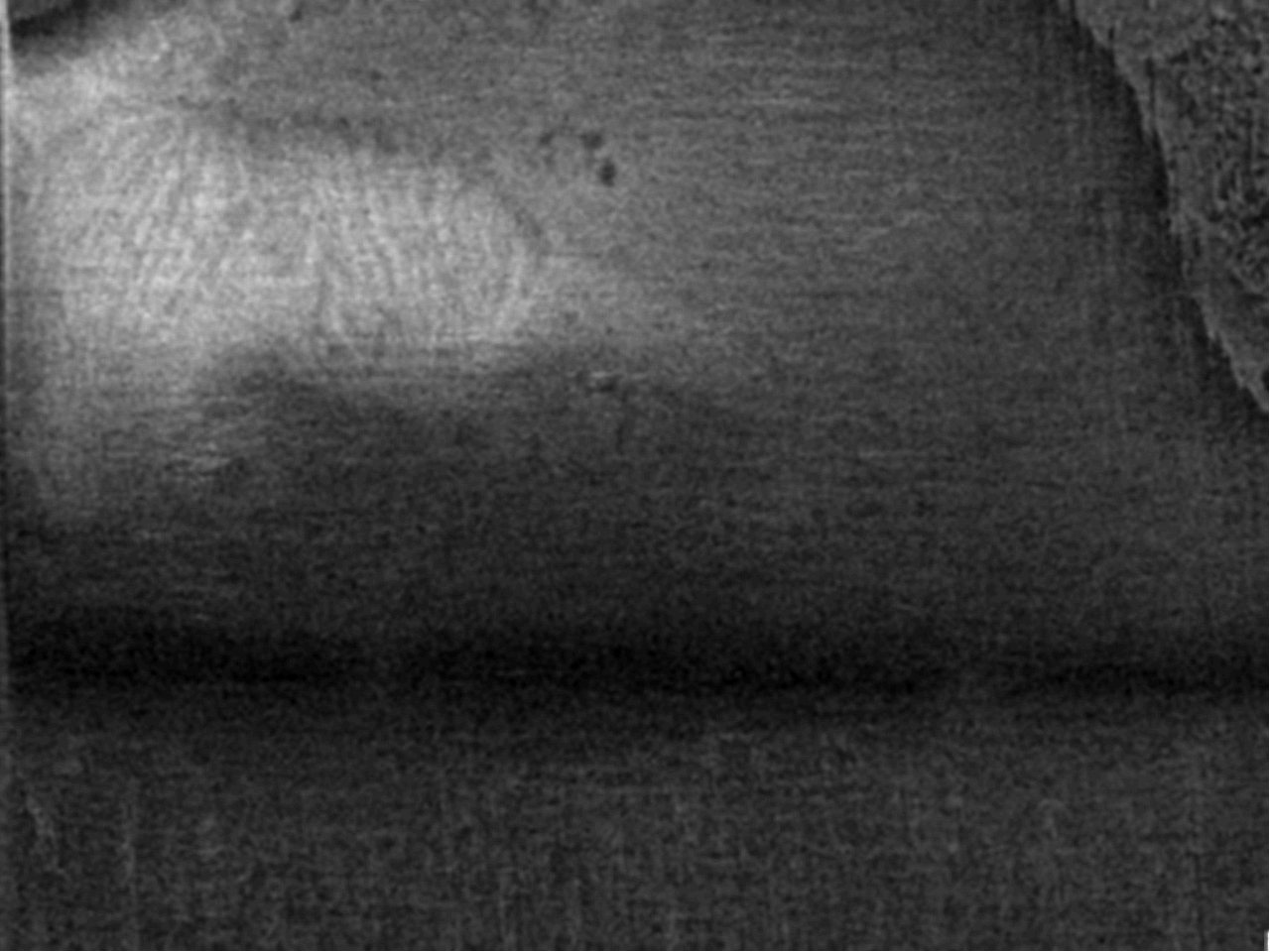}
        \end{minipage}
       
    \end{minipage}

    \caption{Images of \textit{Euglena gracilis} at a variety of sampling percentages, and various processing steps. (Column 1) A single frame from the data cube, without processing. (Column 2) The respective reconstruction of each image displayed in column 1. (Column 3) The integrated data stack from combining all 100 frames of data, without reconstruction. (Column 4) The integrated data stack from combining all 100 reconstructions of the data. Brightness and contrast of images has been modified for viewing.}
    \label{fig:main}
    
\end{figure*}

Fig.~\ref{fig:main} shows various conditions used to image a single prepared face of an algae, \textit{Euglena gracilis}, prepared by FIB-milling. For a series of sampling percentages, a set of 100 images (or frames) was acquired with a dwell time of 1$\mu$s. As can be seen in the first column of Fig.~\ref{fig:main}, as the sampling percentage decreases, by the application of probe subsampling, the image becomes darker as a smaller portion of the data is acquired, with non-sampled data being represented by black pixels. This sampling percentage directly correlates to both the time to acquire the frame and the relative electron dose use to form it - a frame acquired at 10\% sampling was acquired in a tenth the time and electron dose compared to the 100\% sampling equivalent image. 
The second column of Fig.~\ref{fig:main} shows this individual frame reconstructed using a 3-dimensional patch through SenseAI, where the whole stack of 100 frames was included in the dictionary learning process. For all cases, from high sampling to low sampling, this reconstruction visually shows a quality increase over the subsampled frame acquisition.

\begin{figure}
    \centering
    \includegraphics[width=\linewidth]{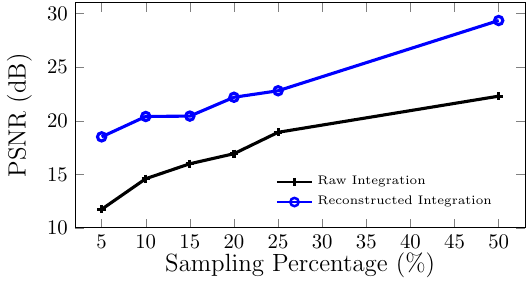}
    \caption{Peak signal-to-noise ratio (PSNR) for the raw integration and reconstructed integration images at various sampling percentages. Each data set is compared to its own 100\% sampling ratio variant. Reconstruction by SenseAI increases image quality by approximately 5.5 dB. } \vspace{-0.5cm}
    \label{fig:psnr}
\end{figure}

Another option for treating subsampled data without requiring a reconstruction is to simply integrate the data, and this is shown in column three of Fig.~\ref{fig:main}, whereby each pixel value is determined by the non-zero mean of the pixel values in the integrating axis. For 100 frames and line hop sampling, this method is valid down to 15\% sampling, where image artefacts are beginning to appear. At 10\% and below, significant distortions are present. In all cases however, benefit is seen through inpainting, as seen in column four: images formed by integrating the entire stack of reconstructions, akin to the regular method of integrating the entire stack of fully sampled images. For each of these integrated images (columns three and four), the peak signal-to-noise ratio, an image quality metric, was calculated. Reconstruction by SenseAI showed an average increase in image quality of 5.5 dB when compared to the equivalent image at 100\% sampling, as shown in Fig.~\ref{fig:psnr}. This can be interpreted as a significant increase in image quality due to inpainting, which enables subsampling as a valid imaging tool to perform fast, low dose SEM imaging. 

\section{Conclusions and Future Work}

In this work, subsampling and inpainting methods have been used to image a cryo-fixed vitrified algae using scanning electron microscopy, enabled by software and hardware supplied by SenseAI and Quantum Detectors. The efficacy of these methods was significantly increased through the application of 3-dimensional dictionary learning and inpainting, wherein an entire data stack was used to inform the data recovery processes. This led to a significant increase in image quality for both single frames and integrated images, enabling both time resolved SEM imaging as well as high-resolution imaging of this beam sensitive material under low-dose conditions. 

The positive results of applying 3-dimensional dictionary learning and inpainting to this integrated-SEM imaging framework demonstrates an important fact: higher dimensionality data provides unique benefits when performing dictionary learning. This is naturally understandable as learning and inpainting efficacy generally increases as the amount of data (and processing power) available to the algorithm increases. 

A natural extension of these methods is to move towards 4-dimensional imaging, wherein FIB-enabled slice-and-view methods inform the fourth dimension, \textit{i.e.,} tomography. This theoretical experiment begins with a data cube of dimension (X, Y, frames, slices), where all four axes are subsampled in some manner. This extension would, in theory, further increase the efficiency of the learning algorithm, and as such increase reconstruction quality and enable even lower sampling rates to be considered. However, a major technological issue stands in the way, in that the size of the data that can be processed by SenseAI is currently limited to the amount of memory on the GPU available. Currently, meaningful 4-dimensional cryo FIB-SEM data volumes exceed SenseAI's capability, and only 3-dimensional data sets such as the one presented in this work are capable for the time being, with work regarding 4-dimensional inpainting reserved for future work. Other ongoing work related to dictionary learning and inpainting of 4D-STEM data using SenseAI corroborates this concept, though the data volumes in 4D-STEM are significantly smaller [\cite{robinson2023simultaneous}]. 

\section{Acknowledgements}

The authors would like to acknowledge JEOL (UK) Ltd and Quantum Detectors for enabling and supporting this work.

\section{Competing Interests}

Authors D. Nicholls, J. Wells, A. Robinson, and N. D. Browning are employed by SenseAI Innovations Ltd. Author D. McGrouther is employed by JEOL (UK) Ltd. All other authors declare no competing interests.


\bibliography{references}
\end{document}

%% file: packages.tex
\usepackage{amssymb}
\usepackage{authblk}
\usepackage{graphicx}
\usepackage[caption=false,font=footnotesize]{subfig}

\usepackage{natbib}
\setcitestyle{authoryear,open={(},close={)}}

\usepackage{stmaryrd}
\usepackage{xcolor}
\usepackage{mathtools}
\usepackage{float}
\usepackage{textcomp}
\usepackage{caption}

\usepackage{geometry}